\documentclass[%
 reprint, footinbib,
 amsmath,amssymb,
 aps, longbibliography
]{revtex4-1}

\usepackage{gensymb}
\usepackage{graphicx}
\usepackage{dcolumn}
\usepackage{bm}
\usepackage{siunitx}
\usepackage[usenames, dvipsnames]{color}
\usepackage{soul}
\usepackage{chngcntr}

\newcommand{\tapp}{\theta_\text{app}}

\sethlcolor{yellow}

\definecolor{darkblue}{rgb}{0,0,0.5}
\setulcolor{darkblue}

\begin{document}

\title{Film dynamics and lubricant depletion by droplets moving on lubricated surfaces}
\author{Michael J. Kreder$^{1}$}
\thanks{These two authors contributed equally}
\author{Dan Daniel$^{1,2}$}
\thanks{These two authors contributed equally}
\author{Adam Tetreault$^{3}$}
\author{Zhenle Cao$^{3}$}
\author{Baptiste Lemaire$^{1}$}
\author{Jaakko V. I. Timonen$^{1,5}$}
\author{Joanna Aizenberg$^{1,3,4}$}%
 \email{jaiz@seas.harvard.edu}
\affiliation{$^{1}$John A. Paulson School of Engineering and Applied Sciences, Harvard University,      
	Cambridge, MA 02138, USA}
\affiliation{$^{2}$Institute for Materials Research and Engineering, 2 Fusionopolis Way, Singapore 138634}
\affiliation{$^{3}$Wyss Institute for Biologically Inspired Engineering, Harvard University, 
    Cambridge, MA 02138, USA} 
\affiliation{$^{4}$Department of Chemistry and Chemical Biology, Harvard University, 
    Cambridge, MA 02138, USA} 
\affiliation{$^{5}$Department of Applied Physics, Aalto University School of Science, 
    Espoo, FI-02150, Finland}

\begin{abstract}

Lubricated surfaces have shown promise in numerous applications where impinging foreign droplets must be removed easily; however, before they can be widely adopted, the problem of lubricant depletion, which eventually leads to decreased performance, must be solved. Despite recent progress, a quantitative mechanistic explanation for lubricant depletion is still lacking. Here, we first explained the shape of a droplet on a lubricated surface by balancing the Laplace pressures across interfaces. We then showed that the lubricant film thicknesses beneath, behind, and wrapping around a moving droplet change dynamically with droplet's speed---analogous to the classical Landau-Levich-Derjaguin problem. The interconnected lubricant dynamics results in the growth of the wetting ridge around the droplet, which is the dominant source of lubricant depletion. We then developed an analytic expression for the maximum amount of lubricant that can be depleted by a single droplet. Counter-intuitively, faster moving droplets subjected to higher driving forces deplete less lubricant than their slower moving counterparts. The insights developed in this work will inform future work and the design of longer-lasting lubricated surfaces. 

\end{abstract}
  
\maketitle
\section{Introduction}

The ability of liquid lubricant on surfaces to reduce \textit{solid-solid} friction has been widely known since antiquity \cite{harris1974lubrication, fall2014sliding}; examples include the ubiquitous use of lubricant oil between the moving parts of a machine and the synovial fluid found naturally in the joint cavities of our bodies \cite{reynolds1886theory, briscoe2006boundary}. The idea of using lubricant to reduce \textit{solid-liquid} friction is relatively new: when infused with suitable lubricants, surfaces can exhibit excellent liquid-repellency \cite{quere2005non, wong_SLIPS_2011,lafuma_slippery_2011}. Such surfaces, known in the literature as Slippery Lubricant Infused Porous Surfaces (SLIPS), also show promise in various applications, including in biomedical devices and anti-ice materials \cite{sunny2016transparent,leslie2014bioinspired, kim_Ice_2012,kreder_design_2016,mistura2017drop}. The origin of repellency in SLIPS is largely due to the presence of a stable lubricant film above the solid surface; however, lubricant can be lost due to various factors (body forces, evaporation/solubility, shear, etc.), eventually leading to decreased performance \cite{wexler2015shear, rykaczewski2013mechanism}. Many strategies have been proposed to retain the lubricant overlayer, ranging from the choice of structures (nanoscale vs. microscale, periodic vs. random, etc.) \cite{kim_hierarchical_2013, kim2016delayed}, to the choice of lubricant (high vs. low viscosity lubricant), and finally to the use of patterned wettability on a surface \cite{wexler2015robust}.  

Despite recent progress, a quantitative mechanistic understanding of lubricant depletion due to a moving droplet has not been reported in the literature. Here, we begin by using geometric arguments and quasi-static approximations---when balancing Laplace pressures across various interfaces---to deduce the shape of a droplet on a lubricated surface. We then proceed to establish scaling arguments for the dynamic behavior of lubricant around a moving droplet, by greatly expanding on the Landau-Levich-Derjaguin (LLD) analysis first outlined in \citeauthor{daniel_oleoplaning_2017} \cite{daniel_oleoplaning_2017}. We validated this model by using white-light interferometry to measure the dynamically-changing lubricant thicknesses behind, underneath, and wrapping around a moving water droplet. 

In our previous work, we showed that the LLD analysis can be used to model droplet mobility on lubricated surfaces \cite{daniel_oleoplaning_2017}. Here, we extend this analysis to directly model lubricant depletion and demonstrate the important role the wetting ridge plays, by showing explicitly that the wetting ridge (and its growth) is the dominant source of depletion.

\section{Results and Discussion}
\subsubsection{Droplet geometry and Laplace pressures on SLIPS}

Recent work by \citeauthor{semprebon2017apparent} and \citeauthor{tress2017shape} used numerical methods to solve the Young-Laplace equation for the droplet geometry on a lubricant-infused surface \cite{semprebon2017apparent, tress2017shape}. Our analysis is consistent with previous work, but we make a number of simplifying assumptions---for the case when the wetting ridge is much smaller than the droplet---that allow for an analytical solution and a simple physical interpretation for the geometry of a millimetric-sized droplet on well-designed SLIPS.

\begin{figure*}
\includegraphics[scale=0.34]{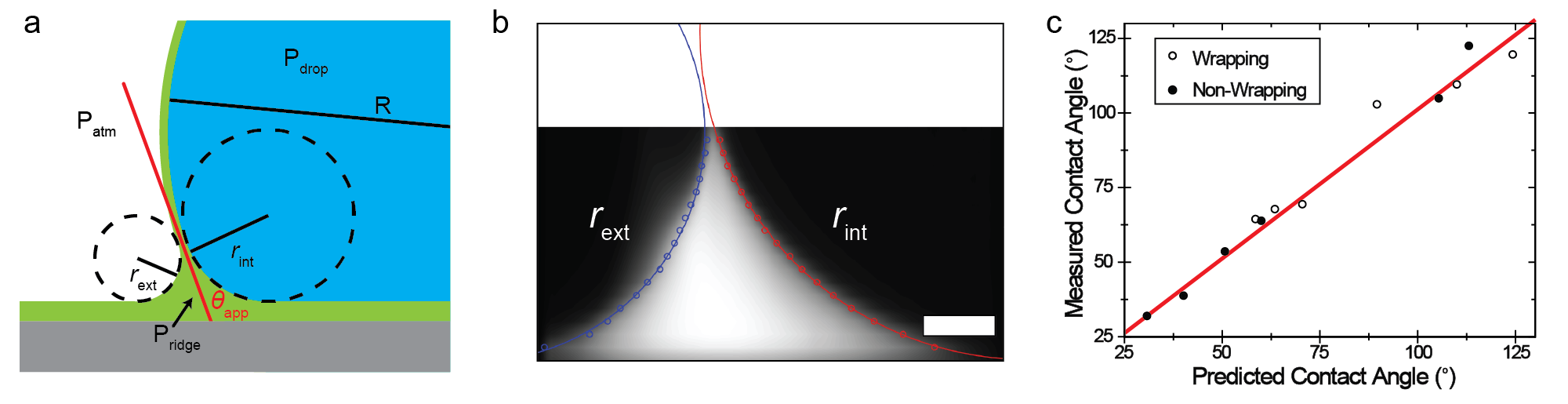} 
\caption{\label{fig:geometry} a) Schematic showing the geometry of a droplet when a wrapping layer is present ($S_{ld}>0$). b) Confocal image confirming the profile of the wetting ridge (scale bar = 25 $\mu$m) around a droplet on fluorescently-dyed silicone oil. c) Measured apparent contact angle vs. predicted apparent contact angle based on Equation \ref{eq:ContactAngle}. The red line indicates a slope of 1.}
\end{figure*}

As shown in Figure \ref{fig:geometry}a, there are three important length-scales to consider: the external radius of the wetting ridge $r_{\text{ext}}$, the internal radius of the wetting ridge $r_{\text{int}}$, and the radius of the droplet itself $R$, whereby $r_{\text{ext}} \sim r_{\text{int}} \ll R$. Note that the micron thicknesses of the lubricant on the substrate outside the droplet, underneath the droplet, and wrapping around the droplet are much thinner than the size of the wetting ridge and do not directly affect the droplet geometry. In our schematic, there is a stable lubricant film underneath the droplet, meaning that there is no well-defined contact angle between the lubricant and the solid \cite{daniel_oleoplaning_2017}. While this is not always the case, the contact angle that the lubricant makes with the solid substrate is close to zero for many well-designed surfaces, even in the absence of a stable intercalating film \cite{schellenberger2015direct}. 

 
We begin by considering a droplet with a wrapping layer of lubricant over it, which occurs when the spreading coefficient of lubricant over the droplet is positive, that is $S_{ld} = \gamma_{dv}-\gamma_{lv}-\gamma_{ld}>0$ where $\gamma_{dv}$, $\gamma_{lv}$, and $\gamma_{ld}$ are the interfacial energies of the droplet-vapor, lubricant-vapor, and lubricant-droplet interfaces, respectively \cite{smith2013droplet, schellenberger2015direct}. The geometry of the sessile droplet, ignoring the effects of gravity, can be understood by equating the Laplace pressures across the different interfaces in the system. The pressure in the drop $P_{\text{drop}}$ can be deduced by applying the Young-Laplace equation across the two interfaces of the wrapping layer, giving
\begin{equation} \label{eq:Pdrop}
P_{\text{drop}} = P_{\text{atm}}+\frac{2 \gamma_{\text{eff}}}{R} = P_{\text{atm}} + \frac{2 (\gamma_{lv} + \gamma_{ld})}{R},  
\end{equation}
while the pressure in the wetting ridge can be deduced from the Laplace pressure either across the air-lubricant or lubricant-droplet interface, giving
\begin{equation} \label{eq:Pridge}
\begin{split}
P_\text{ridge} &= P_\text{atm}-\gamma_{lv} \left(\frac{1}{r_\text{ext}} - \frac{1}{a} \right) \\
&= P_\text{drop}-\gamma_{ld} \left(\frac{1}{r_{\text{int}}} + \frac{1}{a} \right), 
\end{split}
\end{equation}
where $a$ is the base radius of the droplet. 

Comparing Equations \ref{eq:Pdrop} and \ref{eq:Pridge}, and noting that $R\approx a$ for droplets with $\theta_{\text{app}} \approx 90 \degree$, which is true for water droplets on typical SLIPS, we find that  
\begin{equation} \label{eq:CurveBalance}
\frac{\gamma_{ld}}{r_{\text{int}}} = \frac{\gamma_{lv}}{r_{\text{ext}}}+\frac{\gamma_{lv}+\gamma_{ld}}{R},
\end{equation}
where the droplet radius $R$ is set by the volume of the droplet $V$ and the apparent contact angle $\theta_{\text{app}}$ it makes with the surface, i.e. $V = \frac{\pi}{3} R^{3}(2 + \cos \theta_{\text{app}})(1 - \cos \theta_{\text{app}})^{2}$.   

To verify Equation \ref{eq:CurveBalance}, we imaged the wetting ridge using fluorescence confocal microscopy (Figure \ref{fig:geometry}b) \cite{schellenberger2015direct}. We measured $R$, $r_{\text{int}}$, and $r_{\text{ext}}$ for droplets of 3 and 8 $\mu$l, and found good agreement (within 3 \%) between values predicted from Equation \ref{eq:CurveBalance} and experimental values (Supplementary Section S2 and Table S1).

There has been some debate over the correct physical interpretation of $\theta_{\text{app}}$ for SLIPS, which is the angle observed using conventional optical contact-angle instruments \cite{schellenberger2015direct,guan2015evaporation}. Interestingly, a lubricated surface approaches an idealized Young's surface for a vanishingly small wetting ridge, since the there is no contact line pinning for an atomically smooth liquid-liquid interface. Hence, $\tapp$ can be described by a modified Young's equation: 
\begin{equation} \label{eq:ContactAngle}
\cos{\theta_{\text{app}}} = \frac{\gamma_{lv}-\gamma_{ld}}{\gamma_{\text{eff}}},
\end{equation}
where the solid phase is replaced by the lubricant oil (l) phase and $\gamma_{\text{eff}} = \gamma_{lv} + \gamma_{ld}$ or $\gamma_{dv}$ for droplets with and without a wrapping layer, respectively \cite{semprebon2017apparent}. Equation \ref{eq:ContactAngle} can be obtained by either minimizing energy or by balancing forces due to the interfacial tensions at the ridge, similar to argument originally proposed by Young \cite{young1805essay} (Supplementary Figure S1). Alternatively, $\tapp$ can be deduced using purely geometrical considerations (Supplementary Figure S2). As shown in Figure \ref{fig:geometry}c, there is good agreement between experimentally measured contact angles---both by this group \cite{wong_SLIPS_2011} and others \cite{schellenberger2015direct}---and the contact angles predicted by Equation \ref{eq:ContactAngle} (See Supplementary Table S2 for data used in Figure \ref{fig:geometry}c). 

\begin{figure}
\includegraphics[scale = .31]{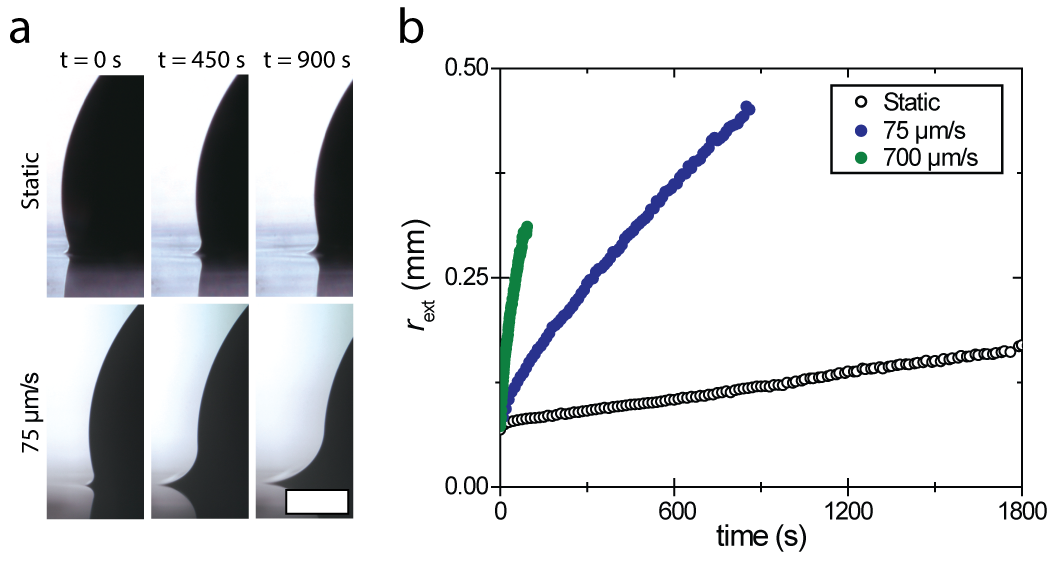}
\caption{\label{fig:ridge} a) Time-lapse images of the wetting ridge for static and moving droplets (scale bar = 0.5 mm). b) Growth of $r_\text{ext}$ over time for static and moving droplets. As can be seen in both (a) and (b), the wetting ridge grows more quickly for the moving droplets. In all cases, 25 $\mu$l droplets were placed on a surface infused with 20 cP silicone oil with an initial film thickness of 4 $\mu$m.}
\end{figure}

Equations \ref{eq:CurveBalance} and \ref{eq:ContactAngle} are true only when $r_{\text{ext}} \ll R$ and $r_{\text{ext}} \ll l_{c}$, where $l_{c} = (\gamma_{lv}/\rho_{l} g)^{1/2} \sim$ mm is the capillary length for lubricant of density $\rho_{l}$. The wetting ridge is a low pressure region and it will grow in size until $r_\text{ext}\sim R$ or $r_\text{ext}\sim l_c$ as lubricant flows from the surrounding area into the ridge, analogous to the flow of liquid from the lamellae into the plateau borders in liquid foams \citep{cantat2013foams}. As the wetting ridge size grows and approaches $l_{c}$, it can no longer be approximated as an arc of a circle with radius $r_{\text{ext}}$, but is described instead by a Bessel function \cite{goodrich1961mathematical, schellenberger2015direct}. In practice, for millimetric droplets on micron-thick lubricant films, the growth of $r_\text{ext}$ around a \textit{static} droplet is limited by thin-film flow and does not approach $l_{c}$ even after a long time. For example, a 25 $\mu$l droplet sitting on 4 $\mu$m thick 20 cP silicone oil had a wetting ridge with an initial $r_\text{ext}\approx 50$ $\mu$m, which grew only to about 150 $\mu$m after 30 minutes (Figure \ref{fig:ridge}). The wetting ridge can, however, grow considerably faster for a \textit{moving} droplet, as we explore more fully in the following section.

\subsubsection{Lubricant Dynamics \label{sec:dynamics}}
\begin{figure*}
\includegraphics[scale=.13]{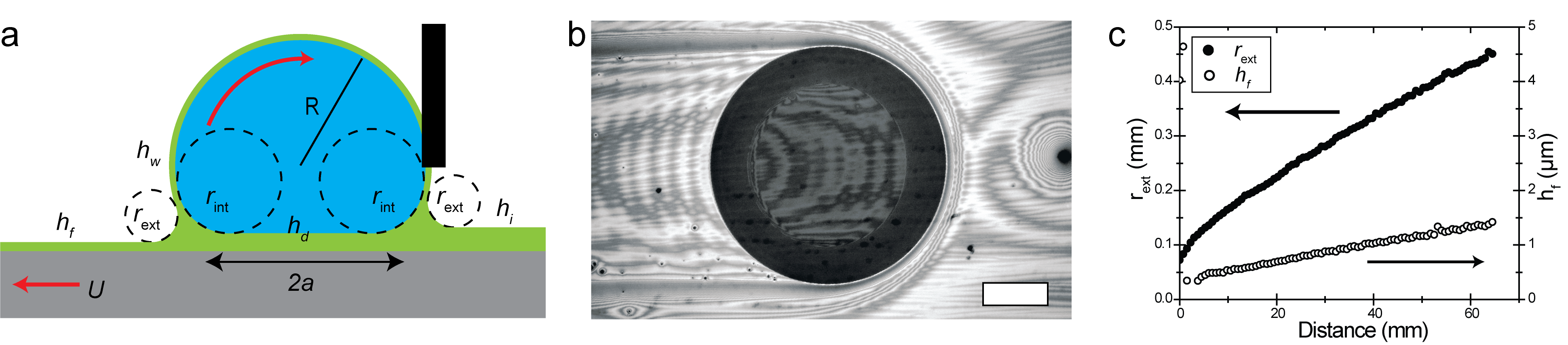} 
\caption{\label{fig:experiment} a) Schematic of the experimental setup used to study the lubricant dynamics. The substrate was moved while the droplet was held in place by a capillary tube, allowing for measurement of the wetting ridge and lubricant thicknesses in various positions.  b) RICM image demonstrating the lubricant profile around a moving droplet (scale bar = 1 mm). See Supplementary Movie S1 for lubricant dynamics visualized using RICM. c) Typical experimental measurement of $h_f$ and $r_{\text{ext}}$ for a droplet moving on a lubricant-infused surface. The first two open circles data refer to initial thickness $h_{i} \approx 4 \mu$m.} 
\end{figure*}

To understand the lubricant dynamics entrained by moving droplets, we tracked the size of the wetting ridge and the thickness of the lubricant in key position with time (Figure \ref{fig:experiment}a). The droplet was held in place by a capillary tube, while the SLIPS sample with initial film thickness $h_{i}$ was moved at controlled speeds $U = 75-700$ $\mu$m/s using a linear motor. In all of our experiments, the SLIPS samples consisted of randomly oriented nano-plates of size 10 nm, spaced 200 nm apart on glass substrates \cite{kim_hierarchical_2013}.

The spatial distribution of the lubricant around and under a moving droplet can be observed using Reflection Interference Contrast Microscopy (RICM) (Figure \ref{fig:experiment}b) \cite{de2015air,daniel_oleoplaning_2017}. Briefly, we shone a monochromatic light of wavelength $\lambda = 532$ nm from beneath a transparent substrate, and in the presence of a thin lubricant film, light reflected off the two film interfaces will interfere either constructively or destructively to form bright or dark fringes, respectively. Between two bright/dark fringes, there is a difference in film thickness of $\lambda/2n_{\text{lub}}$, where $n_{\text{lub}}$ is the refractive index of the lubricant film. The uniformly dark region around the droplet corresponds to the wetting ridge, as light that reflects off the angled ridge is not collected by the objective (Supplementary Section S4).  

The external radius of the wetting ridge $r_{\text{ext}}$, either at the advancing or receding front, was also monitored optically from the side using a high-resolution camera fitted with a microscope objective or a telecentric lens. At the same time, the thicknesses of the initial lubricant film $h_{i}$, at the trail behind the moving droplet $h_f$, under the droplet $h_d$, and wrapping around the droplet $h_w$ were measured using white-light interferometry. White-light reflected off the thin film is collected by an optical fiber of spot size $\sim 50 ~\mu$m and analyzed using a spectrometer. Thicknesses in the range between hundreds of nanometers to tens of microns can be determined this way; details of set-up have been reported in our previous work \cite{daniel_oleoplaning_2017}. In our experiment, $h_{f}$, $h_{d}$, and $h_{i}$ were measured along the midline of the droplet profile, where the lubricant profile is nearly flat with $\Delta h$ of at most $\lambda/4n_{\text{lub}} \sim$ 100 nm.  

\begin{figure*}
\includegraphics[scale=.34]{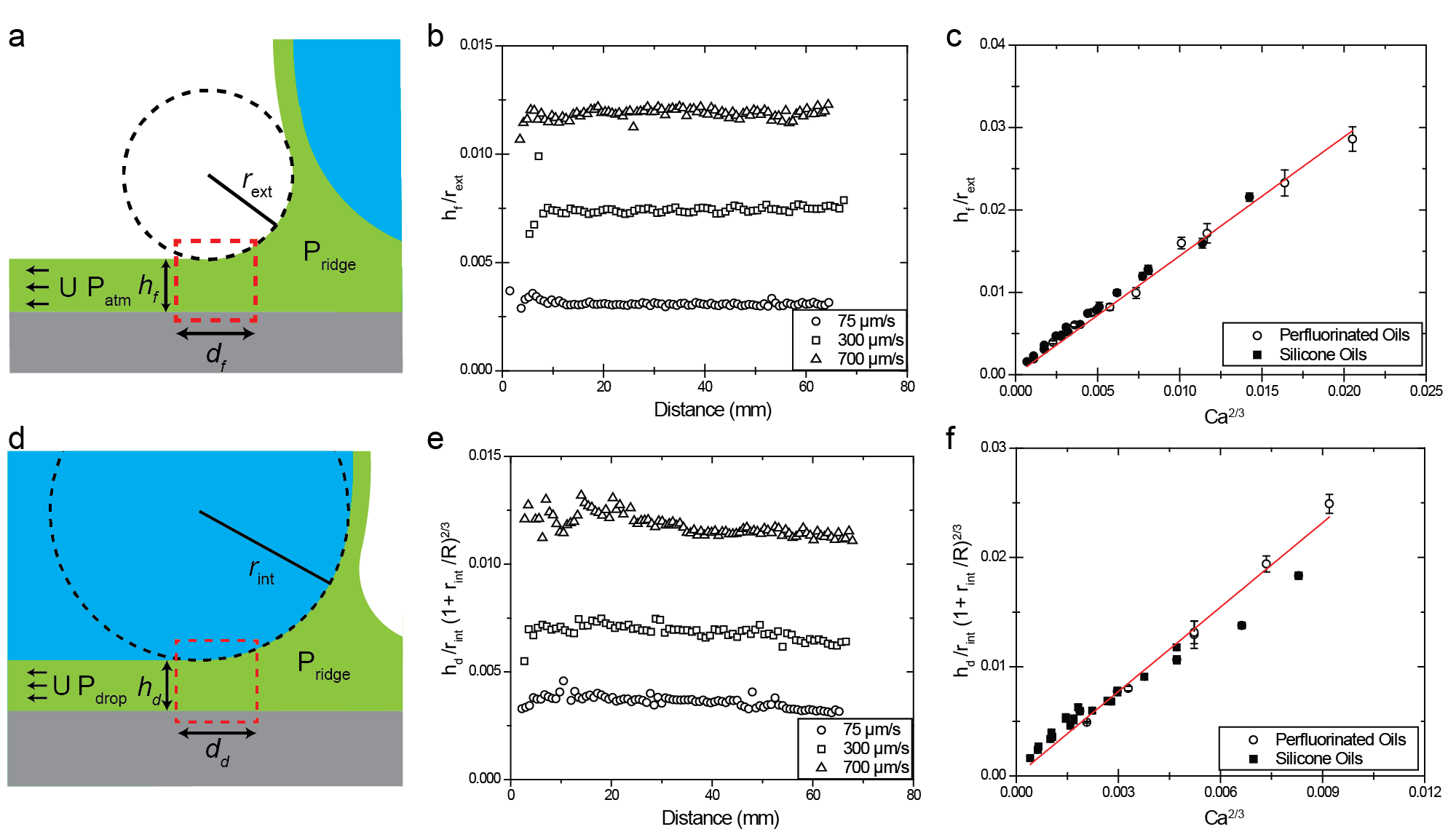} 
\caption{\label{fig:scaling} a) Schematic showing the Landau-Levich-Derjaguin (LLD) film geometry of thickness $h_f$ at the trail behind the moving droplet. b) $h_f/r_{\text{ext}}$ vs distance for droplets moving on silicone oil-infused surfaces under various experimental conditions. c) Scaling of $h_f/r_{\text{ext}}$ with $Ca^{2/3}$ for droplets moving on surfaces infused with silicone and perfluorinated oils of various viscosities. d) Schematic showing the LLD film geometry of thickness $h_d$ beneath the moving droplet. e) $h_d/r_{\text{int}}\; (1+r_{\text{int}}/R)^{2/3}$ vs distance for droplets moving on silicone oil-infused surfaces under various experimental conditions. f) Scaling of $h_d/r_{\text{int}}\; (1+r_{\text{int}}/R)^{2/3}$ with $Ca^{2/3}$ for droplets moving on surfaces infused with silicone and perfluorinated oils of various viscosities. In all cases, the droplet volume was $25$ $\mu$l with initial thickness $h_i = 4$ $\mu$m. Data points in c) and f) are averages of the scaling arguments measured during the entire experiment and error bars indicate the standard deviation.}
\end{figure*}

Experimentally, we found that both $r_{\text{ext}}$ and $h_{f}$ grew (initially) with the distance travelled by the droplet (Figure \ref{fig:experiment}c) as the lubricant was being depleted. We can understand the scaling of $h_{f}$ and $r_{\text{ext}}$, since this behavior is analogous to the classical Landau-Levich-Derjaguin (LLD) problem \cite{derjaguin1943, Landau1942, daniel_oleoplaning_2017, keiser2017drop}. There is a pressure difference between the wetting ridge and the trailing lubricant film behind the droplet that must be balanced by viscous dissipation in the transition region of size $d_f$ (bounded in red in Figure \ref{fig:scaling}a). The film thickness $h_{f}$ can be deduced by balancing $\nabla P$ and $\eta \nabla^{2} U$ in this region, i.e. $(\gamma_{lv}/d_f)(1/r_{\text{ext}}-1/R) \sim \eta U/h_{f}^{2}$, and matching the curvature in this transition region $\partial^2 h_f/\partial^2 x \sim h_f/d_f^2$ with that of wetting ridge $1/r_{\text{ext}}$, i.e. $d_{f} \sim \sqrt{l \;h_{f}}$. Combining these relations gives us the following scaling:
\begin{equation}\label{eq:hfScaling}
h_f/r_{\text{ext}} \; (1-r_\text{ext}/R)^{2/3}\sim Ca_{lv}^{2/3},
\end{equation}
where $Ca_{lv}=\eta U/\gamma_{lv}$ is the corresponding capillary number. 

When $r_{\text{ext}} \ll R$, we recover the classical LLD results where $h_f/r_{\text{ext}} \sim Ca_{lv}^{2/3}$, i.e. $h_f/r_{\text{ext}}$ does not change with distance travelled (Figure \ref{fig:scaling}b). For large droplets ($V = 25$ $\mu$l, $R = 2.05$ mm), the LLD classical law is well-obeyed over a wide range of capillary numbers $Ca_{lv}$ = $10^{-5}-10^{-3}$ with perfluorinated and silicone oils used as lubricants. The red line in Figure \ref{fig:scaling}c shows the best-fit line, with a pre-factor of $\beta = 1.44$, in good agreement with the numerical value obtained in the classical Landau-Levich analysis, $\beta = 0.643(3)^{2/3} \approx 1.34$ \cite{probstein2005physicochemical}.

The data in Figure \ref{fig:scaling}c was taken using $h_i = 4$ $\mu$m and a constant drop volume of $25$ $\mu$l across experiments, but for a range of tested lubricant thicknesses, we found no direct dependence of $h_f/r_{\text{ext}}$ on initial thickness. For droplets smaller than $V < 10 \mu$l, the effect of $R$ can no longer be ignored and Equation \ref{eq:hfScaling} applies. Details on the scaling behavior observed for different initial conditions and droplet radii can be found in the Supplementary Figures S3 and S4.

\begin{figure*}
\includegraphics[scale=0.34]{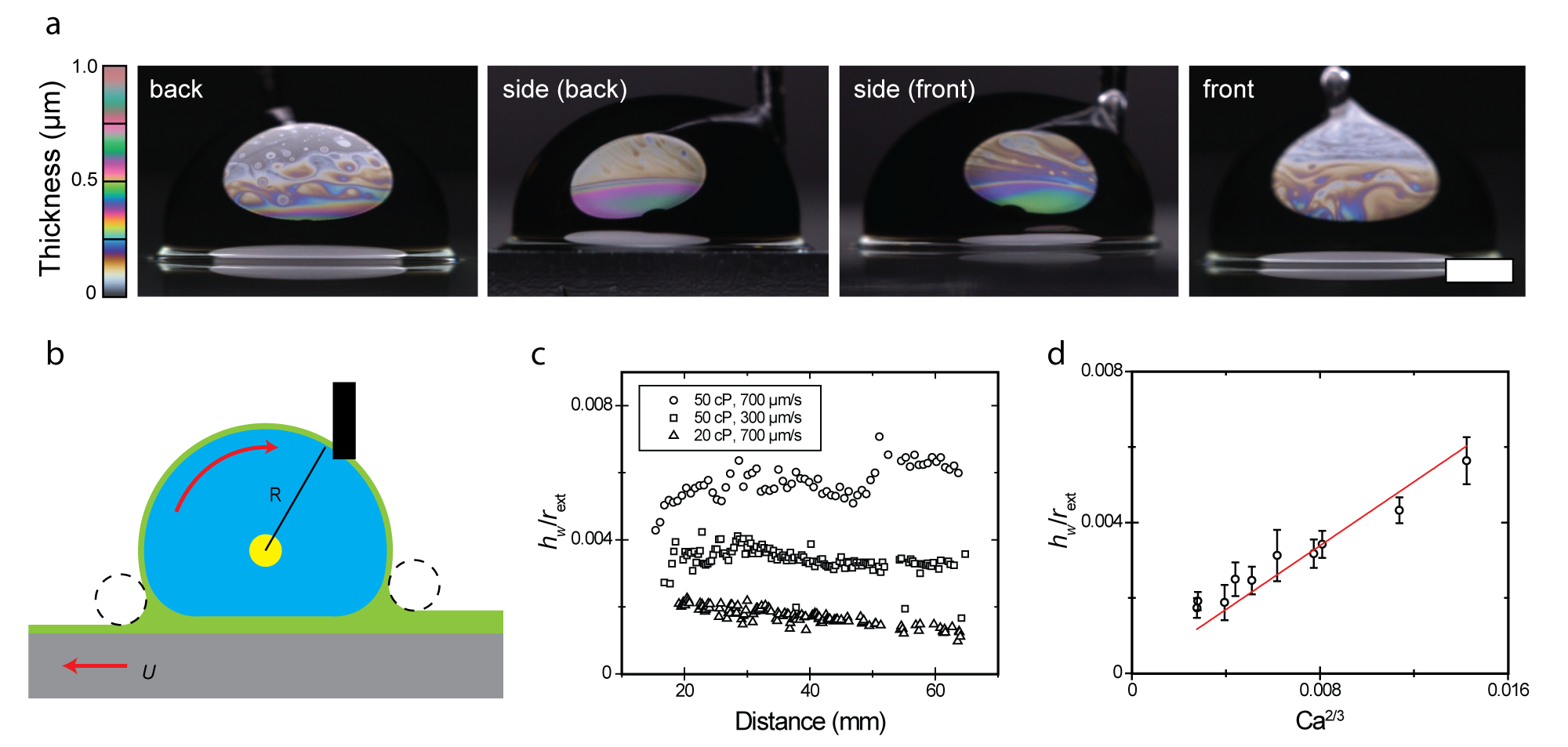}
\caption{\label{fig:Wrap} a) The wrapping layer imaged using color-interferometry at different positions on the surface of a moving droplet. The thickness scale on the left was calculated computationally and should only be used as a guide. $V_{\text{drop}}$ = 25 $\mu$l, $\mu$ = 50 cP, $U$ = 150 $\mu$m/s, $h_i$ = 4 $\mu$m, and L = 40 mm. Scale bar = 1 mm. Note, the ``side (front)" image was reflected to match the orientation of the other images. b) Experimental schematic. The yellow dot indicates the position of the optical probe used to measure $h_w$. c) $h_w/r_{\text{ext}}$ as a function of position for different capillary number experiments. d) Scaling of $h_w/r_{\text{ext}}$ with $Ca^{2/3}$. The line of the best fit is indicated in red.  
}
\end{figure*}

We can make a similar argument about the scaling of $h_d$, the thickness under the droplet, but in this case, the transition region is between the advancing wetting ridge and under the droplet, as shown in Figure \ref{fig:scaling}d. Thus, the pressure difference is $\Delta P = P_{\text{ridge}}-P_{\text{drop}} = -\gamma_{ld}(1/r_{\text{int}}+1/R)$. Using the same arguments we used when deriving Equation \ref{eq:hfScaling}, we arrive at the following result:
\begin{equation}\label{eq:hdScaling}
h_d/r_{\text{int}} \; (1+r_{\text{int}}/R)^{2/3} \sim Ca_{ld}^{2/3},
\end{equation}
where the capillary number here is defined using the interfacial tension between the droplet and the lubricant, i.e. $Ca_{ld}=\eta U/\gamma_{ld}$. As $r_{\text{int}}$ is difficult to measure directly, we measured the external radius of the advancing wetting ridge $r_{\text{ext}}$ and used the relation established in Equation \ref{eq:CurveBalance} to calculate $r_{\text{int}}$. We found that the term $h_d/r_{\text{int}} \; (1+r_{\text{int}}/R)^{2/3}$ is constant throughout the distance travelled in a given experiment (Figure \ref{fig:scaling}e). We see that the scaling behavior follows Equation \ref{eq:hdScaling}, as shown in Figure \ref{fig:scaling}f. The pre-factor in this case is 2.58, which differs substantially from that in classical LLD analysis, since we have ignored the three-dimensional nature of the fluid flow at the droplet base \cite{lhuissier2013levitation}. The deviation between silicone and perfluorinated oils in Figure \ref{fig:scaling}f at high capillary numbers is likely due to the difficulty in aligning the optical probe when the droplet is moving at high speeds.

Note that Equation \ref{eq:hdScaling} is slightly different than the scaling reported by \citeauthor{daniel_oleoplaning_2017}, where the effect of the wetting ridge on the droplet geometry was neglected and it was assumed that $h_d/R \sim Ca_{ld}^{2/3}$ \cite{daniel_oleoplaning_2017}. Importantly, this discrepancy does not change the scaling law for the dissipation force on a moving droplet reported in that paper.   

We expected a similar framework to apply to the dynamic thickness of the lubricant wrapping layer $h_w$. For a static droplet at equilibrium, $h_w$ is stabilized by Van der Waals' interactions and is typically tens of nm thick (Supplementary Figure S5). A moving droplet, however, will entrain a lubricant film with it and $h_w$ thickens with increasing velocity. We follow the analysis used for $h_f$ and $h_d$ to determine the scaling behavior. In this case, the pressure difference in the transition region---between the wetting ridge and the lubricant in the wrapping layer---is $\Delta P = P_{\text{ridge}}-P_{\text{wrapping}} =-\gamma_{lv}(1/r_{\text{ext}}+1/R)$, resulting in the equation:
\begin{equation}\label{eq:hWScaling}
h_w/r_{\text{ext}} \; (1+r_{\text{ext}}/R)^{2/3} \sim Ca_{lv}^{2/3},
\end{equation}
which simplifies to $h_w/r_{\text{ext}} \sim Ca_{lv}^{2/3}$ when the wetting ridge is much smaller than the dimensions of the droplet.

A strict LLD analysis however no longer holds because of the spherical geometry of the droplet. This gives rise to a complex 3D fluid flow and a resulting wrapping layer that is non-uniform in thickness, which can be visualized using color-interferometric techniques \cite{WLI}. We illuminated the droplet using diffuse white lighting; the local film thickness can now be deduced from the ensuing interference patterns captured using a DSLR camera (Figure \ref{fig:Wrap}a). Droplet motion results in a complex lubricant profile. Notably, the lubricant tends to wrap around the sides, forming an extended skirt above the wetting ridge; the lubricant is thicker just above the wetting ridge, but becomes much thinner towards the top of the droplet. Additionally, as compared to the lubricant under and behind a moving droplet, we see much more irregularity in the thickness of the wrapping film, possibly due to the complex interaction between the droplet's internal flow, the lubricant flow, and the draft in ambient air (Supplementary Figure S5 and Movie S2). 

This technique can be used to quantitatively describe the profile of a thin film, since each color corresponds to a specific film thickness \cite{WLI}; it is difficult, however, to distinguish between thicknesses above 500 nm because of the overlap in the color scale (see, for example, the reference colors in Figure \ref{fig:Wrap}a). Hence, to check the validity of equation \ref{eq:hWScaling}, we chose to utilize white-light interference measurements using a spectrometer, as before. The size of the optical probe prevented us from placing it behind the droplet, where the assumptions of LLD are more valid. Hence, we positioned the probe at the side of the droplet, where it is flattest (marked yellow on Figure \ref{fig:Wrap}b). Note that because of the poor refractive index contrast and higher variability in thickness, the minimum $h_w$ that can be accurately measured in this configuration is $\sim 400$ nm, which prevents us from measuring thickness early in an experiment or at lower capillary numbers. 

Because of the complications described above, we do not expect full agreement with LLD results. Experimentally, we find that $h_w/r_{\text{ext}}$ is relatively constant for a given experiment, although there is increased noise due to the variability in $h_w$ (Figure \ref{fig:Wrap}c), and scales with $Ca^{2/3}$ (Figure \ref{fig:Wrap}d). Not surprisingly, however, the experimentally determined pre-factor $\beta$ = 0.42 differs significantly from the classical result of $\beta \approx 1.34$, which was developed for two-dimensional flow. Additionally, we find that there is more noise at high capillary numbers due to more rapid fluctuations in $h_w$ and more challenging probe alignment. A full description of lubricant dynamics in the wrapping layer is a rich and challenging problem, and is outside the scope of this study. 

It has been suggested that the presence of the wrapping layer is the major contribution  to lubricant depletion \cite{smith2013droplet}. In our experiments, the maximum value of $h_w$ measured at the highest capillary number was approximately 1.5 $\mu$m. Applying this value to the total droplet's surface---which would greatly overestimate the amount of lubricant in the wrapping layer---results in a total volume that is about an order of magnitude smaller than the volume of lubricant in the wetting ridge (tens of nl as opposed to hundreds). The wrapping layer is therefore a minor consideration in the overall depletion of the lubricant overlayer; a much more important source of lubricant depletion is the growth of the wetting ridge, which we will discuss in the next section.

\begin{figure*}
\includegraphics[scale=0.34]{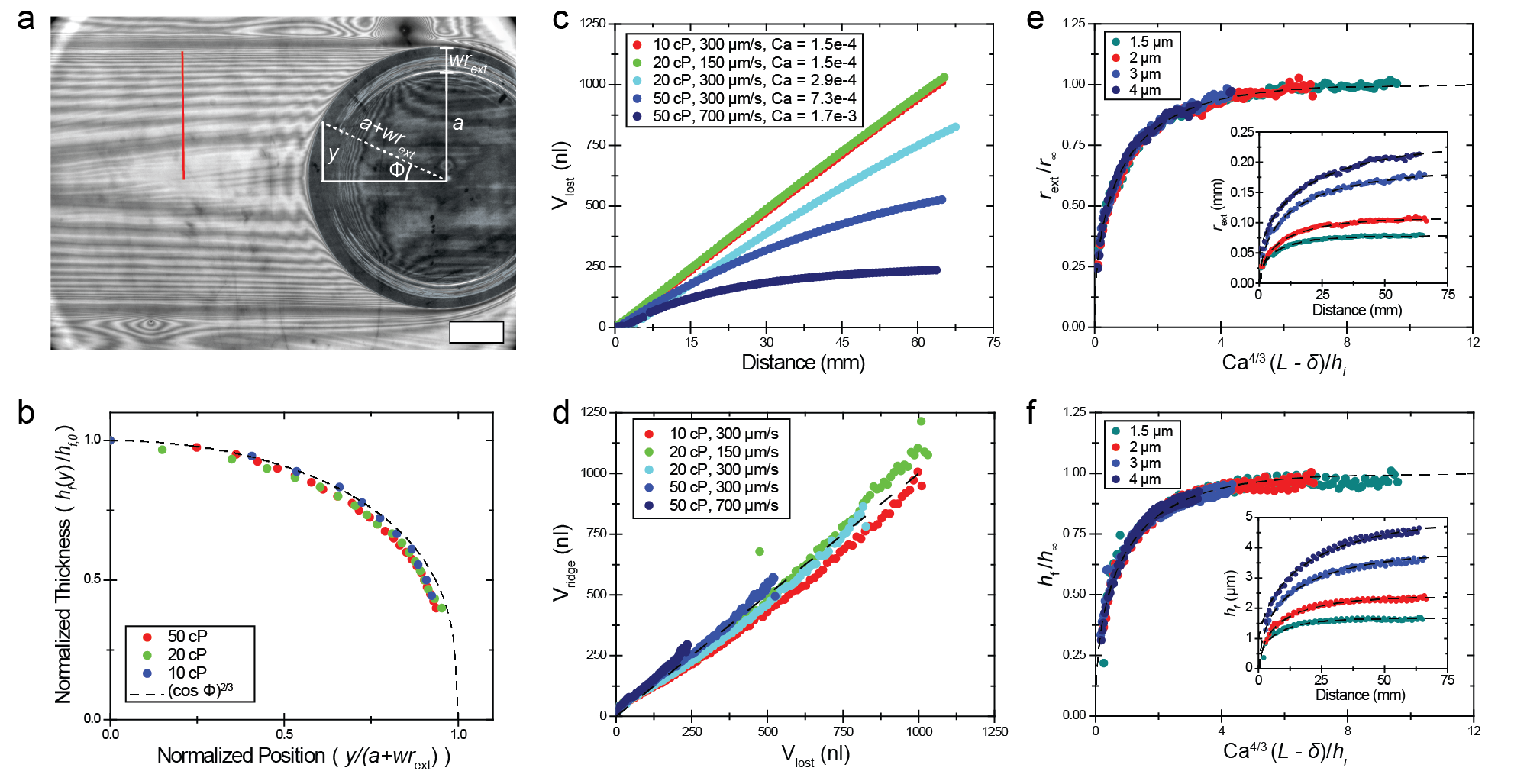}
\caption{\label{fig:depletion} a) RICM of a droplet moving on SLIPS, including the geometric components used to describe the shape of the lubricant behind the droplet (scale bar = 0.5 mm). b) Normalized thickness profiles behind the droplet calculated using RICM intensity profiles (for example, along the red line in (a)) in combination with white light interference measurements. In all cases, $h_i=4$ $\mu$m, $V=25$ $\mu$l, and $U = 300$ $\mu$m/s. c) $V_{\text{lost}}$ vs distance for droplets moving with various capillary numbers over silicone oil SLIPS with $h_i = 4\mu m$. d) $V_{\text{ridge}}$ vs $V_{\text{lost}}$ for experiments on silicone oil with various capillary numbers. The sole fitting parameter, $\alpha = 0.52$. e, f) Non-dimensional plots of the growth of $r_{\text{ext}}$ and $h_{f,0}$ with distance for droplets moving at $700$ $\mu$m/s on 50 cP silicone oil with various initial thicknesses (insets show dimensional quantities). Dashed lines indicate best-fit lines predicted by Equation \ref{eq:L_integral}.
}
\end{figure*}

\subsubsection{Lubricant Depletion \label{sec:Depletion}}

In Section II.2, we have described how the various lubricant dynamics and the resultant film thicknesses ($h_{f}$, $h_{d}$, and $h_{w}$) can be understood in terms of the classical LLD problem. Here, we will explain how lubricant depletion arises from the interconnected lubricant dynamics and how it is intimately linked to the growth of the wetting ridge. In particular, we will explicitly show that the volume of the wetting ridge $V_{\text{ridge}}$ is equal to the volume of lubricant being depleted $V_{\text{lost}}$ due to the change in thicknesses $\Delta h = h_{i}-h_{f}$. With this physical insight, we are then able to describe the process of lubricant depletion in SLIPS fully.    

First, we note that 
\begin{equation} \label{eq:V_ridge}
	V_{\text{ridge}} = \alpha 2\pi a r_{\text{ext}}^2 
\end{equation}
where $\alpha$ is a geometric factor to account for the exact shape. The exact value of $\alpha$ can change as the wetting ridge grows in size and can depend on $Ca$; nevertheless, $\alpha$ should remain at about 0.5, since the wetting ridge can be approximated in the first instance as a triangle. 

To calculate $V_{\text{lost}}$, we need to know the profile of lubricant trail behind a moving droplet, which we deduced using RICM (Figure \ref{fig:depletion}a). Along the midline, we have shown how $h_{f}$ follows the LLD scaling law (Equation \ref{eq:hfScaling}). Away from the midline along $y$, we found that the thickness of the lubricant film $h_{f}(y)$ can be described by a modified LLD scaling law $h_f/r_{\text{ext}} \sim (\eta U_{\phi}/\gamma_{lv})^{2/3}$, where $U_{\phi} = U \cos \phi$ is the radial component of the velocity. Equivalently, $h_f(y)/h_{f,0} = (\cos \phi)^{2/3}$, where $h_{f,0}$ is the film thickness at $y=0$ as described by Equation \ref{eq:hfScaling}.  

Although single wavelength RICM can only give information on the relative lubricant thicknesses at different regions, combining RICM with white-light interferometry measurements allows us to deduce the thickness profile $h_{f}(y)$ unambiguously. For lubricants of different viscosities $\eta$ = 10--50 cP, we found a maximum lubricant thickness at $y=0$, which then decreases monotonically towards the edges of the droplet following the modified LLD scaling law described above (dashed line in Figure \ref{fig:depletion}b). From the profile, we can integrate numerically to calculate the average thickness behind the droplet, resulting in $<h_{f}> \approx 0.85 h_{f,0} = 0.85 \beta r_{\text{ext}} Ca_{lv}^{2/3}$.

Once $r_{\text{ext}}(L)$ and $h_{f,0}(L)$ (and hence $<h_{f}(L)>$) are determined experimentally, the amount of lubricant loss $V_{\text{lost}}(L)$ can then be calculated numerically:
\begin{equation} \label{eq:v_lost}
	V_{\text{lost}}(L) = \int_{0}^{L} 2(a + w r_{\text{ext}}) \Delta h \, dL + V_{o},
\end{equation}
where $L$ is the distance traveled by the droplet, $\Delta h = h_i-<h_{f}>$, $w$ is a geometric factor to account for the added width of the wetting ridge, and $V_{0} = \alpha 2\pi a r_{\text{ext,0}}^{2}$ is volume of the wetting ridge of size $r_{\text{ext,0}}$ created when the droplet was first placed on the surface. 

Figure \ref{fig:depletion}c shows the progression of $V_{\text{lost}}(L)$ for droplets moving with different $Ca_{lv}$ and the same $h_{i}$ = 4 $\mu$m. We note that droplets with higher capillary numbers deplete much less lubricant than slower moving droplets on less viscous lubricants, and $V_{\text{lost}}$ appears to plateau to a maximum value for the highest $Ca_{lv} = 1.7 \times 10^ {-3}$---an observation which we will explain later. We also found that the growth of the wetting ridge $V_{\text{ridge}}(L)$ is directly proportional to $V_{\text{lost}}(L)$, as shown in Figure \ref{fig:depletion}d. The only fitting parameter here is $\alpha_{\text{fit}}=0.52$, close to the expected value in our analysis (Equation \ref{eq:V_ridge}).   

We can also directly predict the lubricant loss $V_{\text{lost}}(L)$ by first solving for $r_{\text{ext}}(L)$, which follows the differential equation $\alpha 4\pi a r_{\text{ext}} (dr_{\text{ext}}/dL) = dV_{lost}/dL = 2(a + w r_{\text{ext}}) \Delta h \approx 2a \Delta h$. Replacing $\Delta h$ with the appropriate scaling law, as discussed earlier, this can be integrated and rearranged to give:
\begin{equation} \label{eq:L_integral}
\ln\bigg(1-\frac{r_\text{ext}}{r_\infty}\bigg) + \frac{r_\text{ext}}{r_\infty} = -\mu_{1} Ca_{lv}^{4/3} \bigg(\frac{L-\delta}{h_i}\bigg),
\end{equation}
where $\mu_{1} = \chi^{2} \beta^{2}/(2 \pi \alpha) \approx 0.5$, $\chi = <h_{f}>/h_{f,0} \approx 0.85$, and $r_\infty = h_i/\chi \beta Ca_{lv}^{2/3}$ is the limiting size of the wetting ridge for a given set of experimental conditions. The integration constant $\delta$ is a small positional offset to account for the experimental error in $L$ and the initial wetting ridge size $r_{\text{ext,0}}$. 

Figure \ref{fig:depletion}e shows the growth of $r_{\text{ext}}(L)$ for droplets moving at 0.7 mm/s on surfaces with different initial lubricant thicknesses $h_{i}$ = 1.5--4 $\mu$m of 50 cP silicone oil. As predicted by Equation \ref{eq:L_integral}, the data collapses to a single curve on a non-dimensional plot of $r_\text{ext}/r_\infty$ against $Ca_{lv}^{4/3} (L-\delta)/h_i$. The dashed lines are the corresponding best-fit curves with $\alpha$, $\delta$, and $\chi$ as fitting parameters. The fitted values are consistent with our previous analysis: $\alpha_\text{fit} = 0.51 \pm 0.05$; $|\delta_\text{fit}|$ is $< 1$ mm, much smaller than the total length travelled $L > 6$ cm, indicating that the error in $L$ and $r_\text{ext,0}$ is minimal; and $\chi_\text{fit} = 0.80 \pm 0.01$, close to the expected value of 0.85. The slight discrepancy in the fitted value of $\chi$ could be due to the finite size of the wetting ridge, which we have ignored when deriving equation \ref{eq:L_integral}. Details on the numerical fit can be found in Supplementary Section S8 and Table S3. 

An analytic solution also exists in the limit that $r_\text{ext}/r_\infty \ll 1$, whereby $\ln(1-r_\text{ext}/r_\infty) + r_\text{ext}/r_\infty \approx -1/2(r/r_\infty)^{2}$, and equation \ref{eq:L_integral} simplifies to   
\begin{equation} \label{eq:small_r}
\begin{split}
r_\text{ext} &= \sqrt{\frac{h_i(L-\delta)}{\pi \alpha}},
\end{split}
\end{equation}
i.e. the growth of the wetting ridge is initially independent of the droplet's speed and fresh drops cause significant local depletion regardless of capillary number. Equation \ref{eq:small_r} can be also be derived by noting that in this limit, $\Delta h \approx h_i$ and hence $dV_{lost}/dL \approx 2a h_i$.

Once $r_{\text{ext}}(L)$ is known, $h_{f,0}(L)$ can be found trivially by applying LLD law (dashed lines in Figure~\ref{fig:depletion}f, where $h_\infty = \beta r_\infty Ca_{lv}^{2/3}$) and $V_{\text{lost}}(L)$ then follows from Equation \ref{eq:v_lost}. Note that as $r_{\text{ext}}$ increases, the Laplace pressure decreases, and the rate of lubricant depletion decreases until it reaches zero when $\Delta h$ = 0 and hence $dV_{\text{lost}}/dL = 0$; this occurs when $h_{f,0} = h_\infty$ and $r_{\text{ext}} = r_\infty$. The maximum amount of lubricant loss can then be deduced by noting that $V_\text{lost,max} = V_{\text{ridge,max}}$, i.e.  
\begin{equation} \label{eq:V_lost_max}
V_{\text{lost,max}} = \alpha 2\pi a r_\infty^{2} = \frac{a \; h_i^2}{\mu_{1}Ca_{lv}^{4/3}}.
\end{equation}
While we have only explicitly demonstrated lubricant depletion due to water droplet, the results presented here, such as Equation \ref{eq:V_lost_max}, can be applied to other liquid droplets (including for small $\theta_{\text{app}}$) as long as $r_{\text{ext}} << R$.

This framework leads us to several conclusions, some of which run against intuitive expectations. Equation \ref{eq:V_lost_max} shows that droplets moving at higher velocities and capillary numbers will cause less lubricant to be lost. From previous work, we know that the drag force for a droplet moving across SLIPS scales with $F_d \sim \gamma R Ca^{2/3}$ \cite{daniel_oleoplaning_2017, keiser2017drop}. Counter-intuitively, this means that droplets moving across the surface with a higher driving force and correspondingly higher shear rate deplete less lubricant than slower-moving droplets. 

We can further consider how the size of the droplet affects lubricant loss. We note that $V_{\text{lost}}$ scales with $a \sim R$, while droplet volume scales with $R^3$. Thus, to minimize depletion, it is advantageous to have one large droplet rather than several smaller droplets of the same total volume. For applications such as condensation where water initially forms on the surface as small, discrete droplets, strategies that promote rapid coalescence into bigger droplets can lead to improved longevity \cite{park2016condensation}.  

In our paper, we have made extensive use of LLD analysis, which is known to break down for large $Ca > 10^{-2}$ \cite{spiers_free_1974, bretherton_motion_1961, white_theory_1965, maleki_landaulevich_2011}. However, we can expect the lubricant thickness to scale as $h \sim r Ca^{\nu}$, where $\nu$ is some exponent which differs from 2/3 \cite{snoeijer2006avoided, maleki_landaulevich_2011, snoeijer2013moving}; in which case, many of the results discussed here can still be applied with some modifications. For example, the amount of lubricant loss can be generalized to $V_{\text{lost,max}} = a \; h_i^2/\mu_{1}Ca_{lv}^{2 \nu}$ (equation \ref{eq:V_lost_max}). Our preliminary results indicate that $\nu \approx 1/4$ for $Ca > 10^{-2}$; its exact value (and its derivation) is still an open question and beyond the scope of this study (Supplementary Figure S6).  

The amount of lubricant lost scales with $h_i^2$. As lubricant is depleted, less and less volume will be lost, with little impact on surface performance since, as we showed in earlier work, $F_{d}$ is independent of $h_i$ \cite{daniel_oleoplaning_2017}. This will remain true until the lubricant thickness becomes nano-metric in size, when dispersion forces (such as van der Waals' interactions) will have to be considered. This will be a natural extension to this work in the future.
 
Finally, we have only used nano-structured surfaces, where the lengthscale of the structures is significantly smaller than the micron-scale lubricant thicknesses. The scaling relationships derived in this work will also apply directly to flat surfaces, such as liquid-infused organogels \cite{urata2015self, cui2015dynamic}, but will have to be modified for microstructured surfaces. However, the main idea outlined in this work---that the wetting ridge is a low-pressure region and its growth drives lubricant depletion---is general and likely to remain true for most surfaces.

\section{Conclusions}

Our work here can be split into three parts. In part I, we show that the geometry of a droplet on SLIPS can be understood by balancing Laplace pressure and using geometric arguments. In part II, we showed how the various lubricant thicknesses change dynamically with speed and can be deduced---analogous to the classical Landau-Levich-Derjaguin problem---by balancing the pressure gradient and viscous stress at the edge of the wetting ridge. Finally, in part III, we use this understanding to identify the growth of the wetting ridge as the main source of lubricant depletion and to quantify the amount of lubricant that a moving droplet collects as it sweeps across a lubricant-infused surface. While we have only explicitly discussed lubricant depletion due to water droplet motion, many of the ideas explored here will be useful in understanding lubricant depletion by other droplets and in very different situations, such as during ice formation and droplet condensation on lubricated surfaces. By identifying the main source and mechanics of lubricant loss on SLIPS, our work will inform the design of longer-lasting lubricant-infused surfaces. 

\section{Materials and Methods \label{sec:mat}} 
\textbf{Boehmite Synthesis.} All SLIPS used in this work were created using transparent thin film of nanostructured boehmite on glass \cite{kim_hierarchical_2013}. Briefly, an alumina sol-gel solution was spin-coated onto 3x1" glass microscope slides at 1000 rpm and dried at 70 \degree C for 1 h. Boehmite was formed from the alumina sol gel by submersion in DI water for 30 min at 60 \degree C. The surfaces were rinsed with DI water and then blown dry with nitrogen. 

\textbf{Surface Functionalization.} For SLIPS infused with a silicone oil as a lubricant, the nanostructured sample was placed in a sealed jar with a small piece of cured Sylgard 184 10:1 PDMS and heated at 235 \degree C for 7 h \cite{PDMSfunc}. Samples were then rinsed with ethanol and dried with nitrogen before application of silicone oil. SLIPS samples infused with perfluorinated polyether oils were first functionalized using perfluoroalkyl phosphate ester (FS100 Surfactant). Boehmite-coated glass slides were submerged in a solution of 95:5:1 by weight ethanol:DI Water:FS100 for 1 h at 70 \degree C. Samples were then rinsed thoroughly with acetone, ethanol, and IPA and blown dry using nitrogen before application of perfluoropolyether oils.  

\textbf{Lubricant Application.} Silicone oils, purchased from Sigma Aldrich, with viscosities in the range of 5 to 50 cP were used. The interfacial tensions (IFTs), as measured using the pendant droplet method, were roughly 19 mN/m in air and 42 mN/m in water, with minor variations due to viscosity. Two perfluoropolyethers were used with viscosities of 23.2 cP and 72.6 cP (Dupont Krytox GPL 100 and 102, respectively). The IFT of the Krytox oils was 16 mN/m in air and 58 mN/m in water, again with minor variation with viscosity. Lubricants were applied by spin-coating at defined speeds and the film thickness was measured using white light interferometry. 

\textbf{Dyed water.} Black, dyed water was used in order to minimize optical reflections during measurements. A thick layer of soot was created on a clean glass petri dish by placing it over a candle-flame. The soot was hydrophilized by exposure to oxygen plasma for 5 min, dissolved in pure de-ionized millipore water, and filtered through a 0.45 $\mu$m cellulose filter. The resulting solution was used as a stock solution that was then diluted 5:1 for all measurements and experiments. The IFT of the dyed droplet in air and with oil was measured using the pendant drop method and found to differ from that of pure water by less than 1\%. 

\textbf{Wetting ridge measurement.} A digital camera (Panasonic GH4) was calibrated for scale and used to take pictures of the wetting ridge profile at a rate of 1 frame per second. The point at which the wetting ridge met the horizontal surface and the point where the wetting ridge met the droplet were tracked in 2D space using an open-source tracking software \cite{tracker}. The wetting ridge radius was calculated as the average of the difference in the $x$ and $y$ coordinates of these two points. In cases where the wetting ridge was discontinuous with the surface of the droplet, the radius was calculated as $(\Delta x^2+\Delta y^2)/2 \Delta y$, where $\Delta x$ and $\Delta y$ are the difference in the $x$ and $y$ coordinates of the two points. 

\textbf{Dynamic Thickness Measurements.} Thickness measurements were performed using an Ocean Optics USB2000+ spectrometer with a halogen lamp as the white-light source. A reflectance-mode optical fiber was placed under the sample and immersion oil was placed between the glass and the probe to eliminate reflection from the back of the glass slide. The spectrum was normalized against the light source, resulting in a series of peaks and valleys appearing due to the difference in path-length between the lubricant-substrate and lubricant-air/water interfaces. By analyzing the wavelengths of the interference maxima and minima, the thickness of the thin films could be unambiguously determined in a range from 400 nm to several microns, as described previously \cite{daniel_oleoplaning_2017}.

\textbf{Reflection Interference Contrast Microscopy.} Samples were imaged with a custom inverted microscope in reflection mode. Monochromatic light was produced by passing broadband LED illumination through a 532 nm filter. Thus, two adjacent maxima or minima differ in thickness by $\lambda/2n_{\text{lub}}$ or 0, and assumptions about the shape/initial thickness must be made in order to obtain a quantitative thickness profile. 

\textbf{Fluorescence Confocal Imaging.} Confocal imaging was done using a Zeiss LSM 700 upright confocal with a 40X water immersion objective. 20 cP silicone oil was dyed with 2.5\% by volume of DFSB-K175 fluorescent dye to generate a fluorescence signal. Further details can be found in Supplementary Section S2. 

\begin{acknowledgments}
The work was supported partially by the ONR MURI Award No. N00014-12-1-0875 and by the Advanced Research Projects Agency-Energy (ARPA-E), U.S. Department of Energy, under Award Number DE-AR0000326. J.V.I.T. was supported by the European Commission through the Seventh Framework Programme (FP7) project DynaSLIPS (project number 626954). We acknowledge the use of the facilities at the Harvard Center for Nanoscale Systems supported by the NSF under Award No. ECS-0335765. M.J.K thanks the Natural Sciences and Engineering Research Council of Canada (NSERC) for a PGS-D scholarship.
\end{acknowledgments}


\begin{thebibliography}{44}%
\makeatletter
\providecommand \@ifxundefined [1]{%
 \@ifx{#1\undefined}
}%
\providecommand \@ifnum [1]{%
 \ifnum #1\expandafter \@firstoftwo
 \else \expandafter \@secondoftwo
 \fi
}%
\providecommand \@ifx [1]{%
 \ifx #1\expandafter \@firstoftwo
 \else \expandafter \@secondoftwo
 \fi
}%
\providecommand \natexlab [1]{#1}%
\providecommand \enquote  [1]{``#1''}%
\providecommand \bibnamefont  [1]{#1}%
\providecommand \bibfnamefont [1]{#1}%
\providecommand \citenamefont [1]{#1}%
\providecommand \href@noop [0]{\@secondoftwo}%
\providecommand \href [0]{\begingroup \@sanitize@url \@href}%
\providecommand \@href[1]{\@@startlink{#1}\@@href}%
\providecommand \@@href[1]{\endgroup#1\@@endlink}%
\providecommand \@sanitize@url [0]{\catcode `\\12\catcode `\$12\catcode
  `\&12\catcode `\#12\catcode `\^12\catcode `\_12\catcode `\%12\relax}%
\providecommand \@@startlink[1]{}%
\providecommand \@@endlink[0]{}%
\providecommand \url  [0]{\begingroup\@sanitize@url \@url }%
\providecommand \@url [1]{\endgroup\@href {#1}{\urlprefix }}%
\providecommand \urlprefix  [0]{URL }%
\providecommand \Eprint [0]{\href }%
\providecommand \doibase [0]{http://dx.doi.org/}%
\providecommand \selectlanguage [0]{\@gobble}%
\providecommand \bibinfo  [0]{\@secondoftwo}%
\providecommand \bibfield  [0]{\@secondoftwo}%
\providecommand \translation [1]{[#1]}%
\providecommand \BibitemOpen [0]{}%
\providecommand \bibitemStop [0]{}%
\providecommand \bibitemNoStop [0]{.\EOS\space}%
\providecommand \EOS [0]{\spacefactor3000\relax}%
\providecommand \BibitemShut  [1]{\csname bibitem#1\endcsname}%
\let\auto@bib@innerbib\@empty
\bibitem [{\citenamefont {Harris}(1974)}]{harris1974lubrication}%
  \BibitemOpen
  \bibfield  {author} {\bibinfo {author} {\bibfnamefont {H.A.}\ \bibnamefont
  {Harris}},\ }\bibfield  {title} {\enquote {\bibinfo {title} {Lubrication in
  antiquity},}\ }\href@noop {} {\bibfield  {journal} {\bibinfo  {journal}
  {Greece and Rome (Second Series)}\ }\textbf {\bibinfo {volume} {21}},\
  \bibinfo {pages} {32--36} (\bibinfo {year} {1974})}\BibitemShut {NoStop}%
\bibitem [{\citenamefont {Fall}\ \emph {et~al.}(2014)\citenamefont {Fall},
  \citenamefont {Weber}, \citenamefont {Pakpour}, \citenamefont {Lenoir},
  \citenamefont {Shahidzadeh}, \citenamefont {Fiscina}, \citenamefont
  {Wagner},\ and\ \citenamefont {Bonn}}]{fall2014sliding}%
  \BibitemOpen
  \bibfield  {author} {\bibinfo {author} {\bibfnamefont {A.}~\bibnamefont
  {Fall}}, \bibinfo {author} {\bibfnamefont {B.}~\bibnamefont {Weber}},
  \bibinfo {author} {\bibfnamefont {M.}~\bibnamefont {Pakpour}}, \bibinfo
  {author} {\bibfnamefont {N.}~\bibnamefont {Lenoir}}, \bibinfo {author}
  {\bibfnamefont {N.}~\bibnamefont {Shahidzadeh}}, \bibinfo {author}
  {\bibfnamefont {J.}~\bibnamefont {Fiscina}}, \bibinfo {author} {\bibfnamefont
  {C.}~\bibnamefont {Wagner}}, \ and\ \bibinfo {author} {\bibfnamefont
  {D.}~\bibnamefont {Bonn}},\ }\bibfield  {title} {\enquote {\bibinfo {title}
  {Sliding friction on wet and dry sand},}\ }\href@noop {} {\bibfield
  {journal} {\bibinfo  {journal} {Phys. Rev. Lett.}\ }\textbf {\bibinfo
  {volume} {112}},\ \bibinfo {pages} {175502} (\bibinfo {year}
  {2014})}\BibitemShut {NoStop}%
\bibitem [{\citenamefont {Reynolds}(1886)}]{reynolds1886theory}%
  \BibitemOpen
  \bibfield  {author} {\bibinfo {author} {\bibfnamefont {O.}~\bibnamefont
  {Reynolds}},\ }\bibfield  {title} {\enquote {\bibinfo {title} {{On the theory
  of lubrication and its application to Mr Beauchamp Tower's experiments,
  including an experimental determination of the viscosity of olive oil}},}\
  }\href@noop {} {\bibfield  {journal} {\bibinfo  {journal} {Proc. R. Soc.
  Lond.}\ }\textbf {\bibinfo {volume} {40}},\ \bibinfo {pages} {191--203}
  (\bibinfo {year} {1886})}\BibitemShut {NoStop}%
\bibitem [{\citenamefont {Briscoe}\ \emph {et~al.}(2006)\citenamefont
  {Briscoe}, \citenamefont {Titmuss}, \citenamefont {Tiberg}, \citenamefont
  {Thomas}, \citenamefont {McGillivray},\ and\ \citenamefont
  {Klein}}]{briscoe2006boundary}%
  \BibitemOpen
  \bibfield  {author} {\bibinfo {author} {\bibfnamefont {W.~H.}\ \bibnamefont
  {Briscoe}}, \bibinfo {author} {\bibfnamefont {S.}~\bibnamefont {Titmuss}},
  \bibinfo {author} {\bibfnamefont {F.}~\bibnamefont {Tiberg}}, \bibinfo
  {author} {\bibfnamefont {R.~K.}\ \bibnamefont {Thomas}}, \bibinfo {author}
  {\bibfnamefont {D.~J.}\ \bibnamefont {McGillivray}}, \ and\ \bibinfo {author}
  {\bibfnamefont {J.}~\bibnamefont {Klein}},\ }\bibfield  {title} {\enquote
  {\bibinfo {title} {Boundary lubrication under water},}\ }\href@noop {}
  {\bibfield  {journal} {\bibinfo  {journal} {Nature (London)}\ }\textbf
  {\bibinfo {volume} {444}},\ \bibinfo {pages} {191--194} (\bibinfo {year}
  {2006})}\BibitemShut {NoStop}%
\bibitem [{\citenamefont {Qu{\'e}r{\'e}}(2005)}]{quere2005non}%
  \BibitemOpen
  \bibfield  {author} {\bibinfo {author} {\bibfnamefont {D.}~\bibnamefont
  {Qu{\'e}r{\'e}}},\ }\bibfield  {title} {\enquote {\bibinfo {title}
  {Non-sticking drops},}\ }\href@noop {} {\bibfield  {journal} {\bibinfo
  {journal} {Rep. Prog. Phys.}\ }\textbf {\bibinfo {volume} {68}},\ \bibinfo
  {pages} {2495} (\bibinfo {year} {2005})}\BibitemShut {NoStop}%
\bibitem [{\citenamefont {Wong}\ \emph {et~al.}(2011)\citenamefont {Wong},
  \citenamefont {Kang}, \citenamefont {Tang}, \citenamefont {Smythe},
  \citenamefont {Hatton}, \citenamefont {Grinthal},\ and\ \citenamefont
  {Aizenberg}}]{wong_SLIPS_2011}%
  \BibitemOpen
  \bibfield  {author} {\bibinfo {author} {\bibfnamefont {T.-S.}\ \bibnamefont
  {Wong}}, \bibinfo {author} {\bibfnamefont {S.~H.}\ \bibnamefont {Kang}},
  \bibinfo {author} {\bibfnamefont {S.~K.~Y.}\ \bibnamefont {Tang}}, \bibinfo
  {author} {\bibfnamefont {E.~J.}\ \bibnamefont {Smythe}}, \bibinfo {author}
  {\bibfnamefont {B.~D.}\ \bibnamefont {Hatton}}, \bibinfo {author}
  {\bibfnamefont {A.}~\bibnamefont {Grinthal}}, \ and\ \bibinfo {author}
  {\bibfnamefont {J.}~\bibnamefont {Aizenberg}},\ }\bibfield  {title} {\enquote
  {\bibinfo {title} {Bioinspired self-repairing slippery surfaces with
  pressure-stable omniphobicity},}\ }\href {\doibase 10.1038/nature10447}
  {\bibfield  {journal} {\bibinfo  {journal} {Nature (London)}\ }\textbf
  {\bibinfo {volume} {477}},\ \bibinfo {pages} {443--447} (\bibinfo {year}
  {2011})}\BibitemShut {NoStop}%
\bibitem [{\citenamefont {Lafuma}\ and\ \citenamefont
  {Qu{\'e}r{\'e}}(2011)}]{lafuma_slippery_2011}%
  \BibitemOpen
  \bibfield  {author} {\bibinfo {author} {\bibfnamefont {A.}~\bibnamefont
  {Lafuma}}\ and\ \bibinfo {author} {\bibfnamefont {D.}~\bibnamefont
  {Qu{\'e}r{\'e}}},\ }\bibfield  {title} {\enquote {\bibinfo {title} {Slippery
  pre-suffused surfaces},}\ }\href {\doibase 10.1209/0295-5075/96/56001}
  {\bibfield  {journal} {\bibinfo  {journal} {Europhys. Lett.}\ }\textbf
  {\bibinfo {volume} {96}},\ \bibinfo {pages} {56001} (\bibinfo {year}
  {2011})}\BibitemShut {NoStop}%
\bibitem [{\citenamefont {Sunny}\ \emph {et~al.}(2016)\citenamefont {Sunny},
  \citenamefont {Cheng}, \citenamefont {Daniel}, \citenamefont {Lo},
  \citenamefont {Ochoa}, \citenamefont {Howell}, \citenamefont {Vogel},
  \citenamefont {Majid},\ and\ \citenamefont
  {Aizenberg}}]{sunny2016transparent}%
  \BibitemOpen
  \bibfield  {author} {\bibinfo {author} {\bibfnamefont {S.}~\bibnamefont
  {Sunny}}, \bibinfo {author} {\bibfnamefont {G.}~\bibnamefont {Cheng}},
  \bibinfo {author} {\bibfnamefont {D.}~\bibnamefont {Daniel}}, \bibinfo
  {author} {\bibfnamefont {P.}~\bibnamefont {Lo}}, \bibinfo {author}
  {\bibfnamefont {S.}~\bibnamefont {Ochoa}}, \bibinfo {author} {\bibfnamefont
  {C.}~\bibnamefont {Howell}}, \bibinfo {author} {\bibfnamefont
  {N.}~\bibnamefont {Vogel}}, \bibinfo {author} {\bibfnamefont
  {A.}~\bibnamefont {Majid}}, \ and\ \bibinfo {author} {\bibfnamefont
  {J.}~\bibnamefont {Aizenberg}},\ }\bibfield  {title} {\enquote {\bibinfo
  {title} {Transparent antifouling material for improved operative field
  visibility in endoscopy},}\ }\href@noop {} {\bibfield  {journal} {\bibinfo
  {journal} {Proc. Natl. Acad. Sci. U.S.A.}\ ,\ \bibinfo {pages} {201605272}}
  (\bibinfo {year} {2016})}\BibitemShut {NoStop}%
\bibitem [{\citenamefont {Leslie}\ \emph {et~al.}(2014)\citenamefont {Leslie},
  \citenamefont {Waterhouse}, \citenamefont {Berthet}, \citenamefont
  {Valentin}, \citenamefont {Watters}, \citenamefont {Jain}, \citenamefont
  {Kim}, \citenamefont {Hatton}, \citenamefont {Nedder}, \citenamefont
  {Donovan} \emph {et~al.}}]{leslie2014bioinspired}%
  \BibitemOpen
  \bibfield  {author} {\bibinfo {author} {\bibfnamefont {D.~C.}\ \bibnamefont
  {Leslie}}, \bibinfo {author} {\bibfnamefont {A.}~\bibnamefont {Waterhouse}},
  \bibinfo {author} {\bibfnamefont {J.~B.}\ \bibnamefont {Berthet}}, \bibinfo
  {author} {\bibfnamefont {T.~M.}\ \bibnamefont {Valentin}}, \bibinfo {author}
  {\bibfnamefont {A.~L.}\ \bibnamefont {Watters}}, \bibinfo {author}
  {\bibfnamefont {A.}~\bibnamefont {Jain}}, \bibinfo {author} {\bibfnamefont
  {P.}~\bibnamefont {Kim}}, \bibinfo {author} {\bibfnamefont {B.~D.}\
  \bibnamefont {Hatton}}, \bibinfo {author} {\bibfnamefont {A.}~\bibnamefont
  {Nedder}}, \bibinfo {author} {\bibfnamefont {K.}~\bibnamefont {Donovan}},
  \emph {et~al.},\ }\bibfield  {title} {\enquote {\bibinfo {title} {A
  bioinspired omniphobic surface coating on medical devices prevents thrombosis
  and biofouling},}\ }\href@noop {} {\bibfield  {journal} {\bibinfo  {journal}
  {Nat. Biotechnol.}\ }\textbf {\bibinfo {volume} {32}},\ \bibinfo {pages}
  {1134--1140} (\bibinfo {year} {2014})}\BibitemShut {NoStop}%
\bibitem [{\citenamefont {Kim}\ \emph {et~al.}(2012)\citenamefont {Kim},
  \citenamefont {Wong}, \citenamefont {Alvarenga}, \citenamefont {Kreder},
  \citenamefont {Adorno-Martinez},\ and\ \citenamefont
  {Aizenberg}}]{kim_Ice_2012}%
  \BibitemOpen
  \bibfield  {author} {\bibinfo {author} {\bibfnamefont {P.}~\bibnamefont
  {Kim}}, \bibinfo {author} {\bibfnamefont {T.-S.}\ \bibnamefont {Wong}},
  \bibinfo {author} {\bibfnamefont {J.}~\bibnamefont {Alvarenga}}, \bibinfo
  {author} {\bibfnamefont {M.~J.}\ \bibnamefont {Kreder}}, \bibinfo {author}
  {\bibfnamefont {W.~E.}\ \bibnamefont {Adorno-Martinez}}, \ and\ \bibinfo
  {author} {\bibfnamefont {J.}~\bibnamefont {Aizenberg}},\ }\bibfield  {title}
  {\enquote {\bibinfo {title} {Liquid-{Infused} {Nanostructured} {Surfaces}
  with {Extreme} {Anti}-{Ice} and {Anti}-{Frost} {Performance}},}\ }\href
  {\doibase 10.1021/nn302310q} {\bibfield  {journal} {\bibinfo  {journal} {ACS
  Nano}\ }\textbf {\bibinfo {volume} {6}},\ \bibinfo {pages} {6569--6577}
  (\bibinfo {year} {2012})}\BibitemShut {NoStop}%
\bibitem [{\citenamefont {Kreder}\ \emph {et~al.}(2016)\citenamefont {Kreder},
  \citenamefont {Alvarenga}, \citenamefont {Kim},\ and\ \citenamefont
  {Aizenberg}}]{kreder_design_2016}%
  \BibitemOpen
  \bibfield  {author} {\bibinfo {author} {\bibfnamefont {M.~J.}\ \bibnamefont
  {Kreder}}, \bibinfo {author} {\bibfnamefont {J.}~\bibnamefont {Alvarenga}},
  \bibinfo {author} {\bibfnamefont {P.}~\bibnamefont {Kim}}, \ and\ \bibinfo
  {author} {\bibfnamefont {J.}~\bibnamefont {Aizenberg}},\ }\bibfield  {title}
  {\enquote {\bibinfo {title} {Design of anti-icing surfaces: smooth, textured
  or slippery?}}\ }\href {\doibase 10.1038/natrevmats.2015.3} {\bibfield
  {journal} {\bibinfo  {journal} {Nat. Rev. Mater.}\ }\textbf {\bibinfo
  {volume} {1}},\ \bibinfo {pages} {15003} (\bibinfo {year}
  {2016})}\BibitemShut {NoStop}%
\bibitem [{\citenamefont {Mistura}\ and\ \citenamefont
  {Pierno}(2017)}]{mistura2017drop}%
  \BibitemOpen
  \bibfield  {author} {\bibinfo {author} {\bibfnamefont {G.}~\bibnamefont
  {Mistura}}\ and\ \bibinfo {author} {\bibfnamefont {M.}~\bibnamefont
  {Pierno}},\ }\bibfield  {title} {\enquote {\bibinfo {title} {Drop mobility on
  chemically heterogeneous and lubricant-impregnated surfaces},}\ }\href@noop
  {} {\bibfield  {journal} {\bibinfo  {journal} {Adv. Phys. X}\ }\textbf
  {\bibinfo {volume} {2}},\ \bibinfo {pages} {591--607} (\bibinfo {year}
  {2017})}\BibitemShut {NoStop}%
\bibitem [{\citenamefont {Wexler}\ \emph
  {et~al.}(2015{\natexlab{a}})\citenamefont {Wexler}, \citenamefont {Jacobi},\
  and\ \citenamefont {Stone}}]{wexler2015shear}%
  \BibitemOpen
  \bibfield  {author} {\bibinfo {author} {\bibfnamefont {J.~S.}\ \bibnamefont
  {Wexler}}, \bibinfo {author} {\bibfnamefont {I.}~\bibnamefont {Jacobi}}, \
  and\ \bibinfo {author} {\bibfnamefont {H.~A.}\ \bibnamefont {Stone}},\
  }\bibfield  {title} {\enquote {\bibinfo {title} {Shear-driven failure of
  liquid-infused surfaces},}\ }\href@noop {} {\bibfield  {journal} {\bibinfo
  {journal} {Phys. Rev. Lett.}\ }\textbf {\bibinfo {volume} {114}},\ \bibinfo
  {pages} {168301} (\bibinfo {year} {2015}{\natexlab{a}})}\BibitemShut
  {NoStop}%
\bibitem [{\citenamefont {Rykaczewski}\ \emph {et~al.}(2013)\citenamefont
  {Rykaczewski}, \citenamefont {Anand}, \citenamefont {Subramanyam},\ and\
  \citenamefont {Varanasi}}]{rykaczewski2013mechanism}%
  \BibitemOpen
  \bibfield  {author} {\bibinfo {author} {\bibfnamefont {K.}~\bibnamefont
  {Rykaczewski}}, \bibinfo {author} {\bibfnamefont {S.}~\bibnamefont {Anand}},
  \bibinfo {author} {\bibfnamefont {S.~B.}\ \bibnamefont {Subramanyam}}, \ and\
  \bibinfo {author} {\bibfnamefont {K.~K.}\ \bibnamefont {Varanasi}},\
  }\bibfield  {title} {\enquote {\bibinfo {title} {Mechanism of frost formation
  on lubricant-impregnated surfaces},}\ }\href@noop {} {\bibfield  {journal}
  {\bibinfo  {journal} {Langmuir}\ }\textbf {\bibinfo {volume} {29}},\ \bibinfo
  {pages} {5230--5238} (\bibinfo {year} {2013})}\BibitemShut {NoStop}%
\bibitem [{\citenamefont {Kim}\ \emph {et~al.}(2013)\citenamefont {Kim},
  \citenamefont {Kreder}, \citenamefont {Alvarenga},\ and\ \citenamefont
  {Aizenberg}}]{kim_hierarchical_2013}%
  \BibitemOpen
  \bibfield  {author} {\bibinfo {author} {\bibfnamefont {P.}~\bibnamefont
  {Kim}}, \bibinfo {author} {\bibfnamefont {M.~J.}\ \bibnamefont {Kreder}},
  \bibinfo {author} {\bibfnamefont {J.}~\bibnamefont {Alvarenga}}, \ and\
  \bibinfo {author} {\bibfnamefont {J.}~\bibnamefont {Aizenberg}},\ }\bibfield
  {title} {\enquote {\bibinfo {title} {Hierarchical or not? {Effect} of the
  length scale and hierarchy of the surface roughness on omniphobicity of
  lubricant-infused substrates},}\ }\href {\doibase 10.1021/nl4003969}
  {\bibfield  {journal} {\bibinfo  {journal} {Nano Lett.}\ }\textbf {\bibinfo
  {volume} {13}},\ \bibinfo {pages} {1793--1799} (\bibinfo {year}
  {2013})}\BibitemShut {NoStop}%
\bibitem [{\citenamefont {Kim}\ and\ \citenamefont
  {Rothstein}(2016)}]{kim2016delayed}%
  \BibitemOpen
  \bibfield  {author} {\bibinfo {author} {\bibfnamefont {J.-H.}\ \bibnamefont
  {Kim}}\ and\ \bibinfo {author} {\bibfnamefont {J.~P.}\ \bibnamefont
  {Rothstein}},\ }\bibfield  {title} {\enquote {\bibinfo {title} {Delayed
  lubricant depletion on liquid-infused randomly rough surfaces},}\ }\href@noop
  {} {\bibfield  {journal} {\bibinfo  {journal} {Exp. Fluids}\ }\textbf
  {\bibinfo {volume} {57}},\ \bibinfo {pages} {81} (\bibinfo {year}
  {2016})}\BibitemShut {NoStop}%
\bibitem [{\citenamefont {Wexler}\ \emph
  {et~al.}(2015{\natexlab{b}})\citenamefont {Wexler}, \citenamefont
  {Grosskopf}, \citenamefont {Chow}, \citenamefont {Fan}, \citenamefont
  {Jacobi},\ and\ \citenamefont {Stone}}]{wexler2015robust}%
  \BibitemOpen
  \bibfield  {author} {\bibinfo {author} {\bibfnamefont {J.~S.}\ \bibnamefont
  {Wexler}}, \bibinfo {author} {\bibfnamefont {A.}~\bibnamefont {Grosskopf}},
  \bibinfo {author} {\bibfnamefont {M.}~\bibnamefont {Chow}}, \bibinfo {author}
  {\bibfnamefont {Y.}~\bibnamefont {Fan}}, \bibinfo {author} {\bibfnamefont
  {I.}~\bibnamefont {Jacobi}}, \ and\ \bibinfo {author} {\bibfnamefont {H.~A.}\
  \bibnamefont {Stone}},\ }\bibfield  {title} {\enquote {\bibinfo {title}
  {Robust liquid-infused surfaces through patterned wettability},}\ }\href@noop
  {} {\bibfield  {journal} {\bibinfo  {journal} {Soft Matter}\ }\textbf
  {\bibinfo {volume} {11}},\ \bibinfo {pages} {5023--5029} (\bibinfo {year}
  {2015}{\natexlab{b}})}\BibitemShut {NoStop}%
\bibitem [{\citenamefont {Daniel}\ \emph {et~al.}(2017)\citenamefont {Daniel},
  \citenamefont {Timonen}, \citenamefont {Li}, \citenamefont {Velling},\ and\
  \citenamefont {Aizenberg}}]{daniel_oleoplaning_2017}%
  \BibitemOpen
  \bibfield  {author} {\bibinfo {author} {\bibfnamefont {D.}~\bibnamefont
  {Daniel}}, \bibinfo {author} {\bibfnamefont {J.V.I.}\ \bibnamefont
  {Timonen}}, \bibinfo {author} {\bibfnamefont {R.}~\bibnamefont {Li}},
  \bibinfo {author} {\bibfnamefont {S.J.}\ \bibnamefont {Velling}}, \ and\
  \bibinfo {author} {\bibfnamefont {J.}~\bibnamefont {Aizenberg}},\ }\bibfield
  {title} {\enquote {\bibinfo {title} {Oleoplaning droplets on lubricated
  surfaces},}\ }\href@noop {} {\bibfield  {journal} {\bibinfo  {journal} {Nat.
  Phys.}\ }\textbf {\bibinfo {volume} {13}},\ \bibinfo {pages} {1020--1025}
  (\bibinfo {year} {2017})}\BibitemShut {NoStop}%
\bibitem [{\citenamefont {Semprebon}\ \emph {et~al.}(2017)\citenamefont
  {Semprebon}, \citenamefont {McHale},\ and\ \citenamefont
  {Kusumaatmaja}}]{semprebon2017apparent}%
  \BibitemOpen
  \bibfield  {author} {\bibinfo {author} {\bibfnamefont {C.}~\bibnamefont
  {Semprebon}}, \bibinfo {author} {\bibfnamefont {G.}~\bibnamefont {McHale}}, \
  and\ \bibinfo {author} {\bibfnamefont {H.}~\bibnamefont {Kusumaatmaja}},\
  }\bibfield  {title} {\enquote {\bibinfo {title} {Apparent contact angle and
  contact angle hysteresis on liquid infused surfaces},}\ }\href@noop {}
  {\bibfield  {journal} {\bibinfo  {journal} {Soft Matter}\ }\textbf {\bibinfo
  {volume} {13}},\ \bibinfo {pages} {101--110} (\bibinfo {year}
  {2017})}\BibitemShut {NoStop}%
\bibitem [{\citenamefont {Tress}\ \emph {et~al.}(2017)\citenamefont {Tress},
  \citenamefont {Karpitschka}, \citenamefont {Papadopoulos}, \citenamefont
  {Snoeijer}, \citenamefont {Vollmer},\ and\ \citenamefont
  {Butt}}]{tress2017shape}%
  \BibitemOpen
  \bibfield  {author} {\bibinfo {author} {\bibfnamefont {M.}~\bibnamefont
  {Tress}}, \bibinfo {author} {\bibfnamefont {S.}~\bibnamefont {Karpitschka}},
  \bibinfo {author} {\bibfnamefont {P.}~\bibnamefont {Papadopoulos}}, \bibinfo
  {author} {\bibfnamefont {J.~H.}\ \bibnamefont {Snoeijer}}, \bibinfo {author}
  {\bibfnamefont {D.}~\bibnamefont {Vollmer}}, \ and\ \bibinfo {author}
  {\bibfnamefont {H.-J.}\ \bibnamefont {Butt}},\ }\bibfield  {title} {\enquote
  {\bibinfo {title} {Shape of a sessile drop on a flat surface covered with a
  liquid film},}\ }\href@noop {} {\bibfield  {journal} {\bibinfo  {journal}
  {Soft Matter}\ }\textbf {\bibinfo {volume} {13}},\ \bibinfo {pages}
  {3760--3767} (\bibinfo {year} {2017})}\BibitemShut {NoStop}%
\bibitem [{\citenamefont {Schellenberger}\ \emph {et~al.}(2015)\citenamefont
  {Schellenberger}, \citenamefont {Xie}, \citenamefont {Encinas}, \citenamefont
  {Hardy}, \citenamefont {Klapper}, \citenamefont {Papadopoulos},\ and\
  \citenamefont {Butt}}]{schellenberger2015direct}%
  \BibitemOpen
  \bibfield  {author} {\bibinfo {author} {\bibfnamefont {F.}~\bibnamefont
  {Schellenberger}}, \bibinfo {author} {\bibfnamefont {J.}~\bibnamefont {Xie}},
  \bibinfo {author} {\bibfnamefont {N.}~\bibnamefont {Encinas}}, \bibinfo
  {author} {\bibfnamefont {A.}~\bibnamefont {Hardy}}, \bibinfo {author}
  {\bibfnamefont {M.}~\bibnamefont {Klapper}}, \bibinfo {author} {\bibfnamefont
  {P.}~\bibnamefont {Papadopoulos}}, \ and\ \bibinfo {author} {\bibfnamefont
  {D.}~\bibnamefont {Butt}, \bibfnamefont {H.-J.and~Vollmer}},\ }\bibfield
  {title} {\enquote {\bibinfo {title} {Direct observation of drops on slippery
  lubricant-infused surfaces},}\ }\href {\doibase 10.1039/C5SM01809A}
  {\bibfield  {journal} {\bibinfo  {journal} {Soft Matter}\ }\textbf {\bibinfo
  {volume} {11}},\ \bibinfo {pages} {7617--7626} (\bibinfo {year}
  {2015})}\BibitemShut {NoStop}%
\bibitem [{\citenamefont {Smith}\ \emph {et~al.}(2013)\citenamefont {Smith},
  \citenamefont {Dhiman}, \citenamefont {Anand}, \citenamefont {Reza-Garduno},
  \citenamefont {Cohen}, \citenamefont {McKinley},\ and\ \citenamefont
  {Varanasi}}]{smith2013droplet}%
  \BibitemOpen
  \bibfield  {author} {\bibinfo {author} {\bibfnamefont {J.~D.}\ \bibnamefont
  {Smith}}, \bibinfo {author} {\bibfnamefont {R.}~\bibnamefont {Dhiman}},
  \bibinfo {author} {\bibfnamefont {S.}~\bibnamefont {Anand}}, \bibinfo
  {author} {\bibfnamefont {E.}~\bibnamefont {Reza-Garduno}}, \bibinfo {author}
  {\bibfnamefont {R.~E.}\ \bibnamefont {Cohen}}, \bibinfo {author}
  {\bibfnamefont {G.~H.}\ \bibnamefont {McKinley}}, \ and\ \bibinfo {author}
  {\bibfnamefont {K.~K.}\ \bibnamefont {Varanasi}},\ }\bibfield  {title}
  {\enquote {\bibinfo {title} {Droplet mobility on lubricant-impregnated
  surfaces},}\ }\href@noop {} {\bibfield  {journal} {\bibinfo  {journal} {Soft
  Matter}\ }\textbf {\bibinfo {volume} {9}},\ \bibinfo {pages} {1772--1780}
  (\bibinfo {year} {2013})}\BibitemShut {NoStop}%
\bibitem [{\citenamefont {Guan}\ \emph {et~al.}(2015)\citenamefont {Guan},
  \citenamefont {Wells}, \citenamefont {Xu}, \citenamefont {McHale},
  \citenamefont {Wood}, \citenamefont {Martin},\ and\ \citenamefont
  {Stuart-Cole}}]{guan2015evaporation}%
  \BibitemOpen
  \bibfield  {author} {\bibinfo {author} {\bibfnamefont {J.~H.}\ \bibnamefont
  {Guan}}, \bibinfo {author} {\bibfnamefont {G.~G.}\ \bibnamefont {Wells}},
  \bibinfo {author} {\bibfnamefont {B.}~\bibnamefont {Xu}}, \bibinfo {author}
  {\bibfnamefont {G.}~\bibnamefont {McHale}}, \bibinfo {author} {\bibfnamefont
  {D.}~\bibnamefont {Wood}}, \bibinfo {author} {\bibfnamefont {J.}~\bibnamefont
  {Martin}}, \ and\ \bibinfo {author} {\bibfnamefont {S.}~\bibnamefont
  {Stuart-Cole}},\ }\bibfield  {title} {\enquote {\bibinfo {title} {Evaporation
  of sessile droplets on slippery liquid-infused porous surfaces (slips)},}\
  }\href@noop {} {\bibfield  {journal} {\bibinfo  {journal} {Langmuir}\
  }\textbf {\bibinfo {volume} {31}},\ \bibinfo {pages} {11781--11789} (\bibinfo
  {year} {2015})}\BibitemShut {NoStop}%
\bibitem [{\citenamefont {Young}(1805)}]{young1805essay}%
  \BibitemOpen
  \bibfield  {author} {\bibinfo {author} {\bibfnamefont {T.}~\bibnamefont
  {Young}},\ }\bibfield  {title} {\enquote {\bibinfo {title} {An essay on the
  cohesion of fluids},}\ }\href@noop {} {\bibfield  {journal} {\bibinfo
  {journal} {Philos. Trans. R. Soc. London}\ }\textbf {\bibinfo {volume}
  {95}},\ \bibinfo {pages} {65--87} (\bibinfo {year} {1805})}\BibitemShut
  {NoStop}%
\bibitem [{\citenamefont {Cantat}\ \emph {et~al.}(2013)\citenamefont {Cantat},
  \citenamefont {Cohen-Addad}, \citenamefont {Elias}, \citenamefont {Graner},
  \citenamefont {H{\"o}hler}, \citenamefont {Pitois}, \citenamefont {Rouyer},\
  and\ \citenamefont {Saint-Jalmes}}]{cantat2013foams}%
  \BibitemOpen
  \bibfield  {author} {\bibinfo {author} {\bibfnamefont {I.}~\bibnamefont
  {Cantat}}, \bibinfo {author} {\bibfnamefont {S.}~\bibnamefont {Cohen-Addad}},
  \bibinfo {author} {\bibfnamefont {F.}~\bibnamefont {Elias}}, \bibinfo
  {author} {\bibfnamefont {F.}~\bibnamefont {Graner}}, \bibinfo {author}
  {\bibfnamefont {R.}~\bibnamefont {H{\"o}hler}}, \bibinfo {author}
  {\bibfnamefont {O.}~\bibnamefont {Pitois}}, \bibinfo {author} {\bibfnamefont
  {F.}~\bibnamefont {Rouyer}}, \ and\ \bibinfo {author} {\bibfnamefont
  {A.}~\bibnamefont {Saint-Jalmes}},\ }\href@noop {} {\emph {\bibinfo {title}
  {Foams: structure and dynamics}}}\ (\bibinfo  {publisher} {OUP Oxford},\
  \bibinfo {year} {2013})\BibitemShut {NoStop}%
\bibitem [{\citenamefont {Goodrich}(1961)}]{goodrich1961mathematical}%
  \BibitemOpen
  \bibfield  {author} {\bibinfo {author} {\bibfnamefont {F.~C.}\ \bibnamefont
  {Goodrich}},\ }\bibfield  {title} {\enquote {\bibinfo {title} {The
  mathematical theory of capillarity. i},}\ }\href@noop {} {\bibfield
  {journal} {\bibinfo  {journal} {Proc. R. Soc. London, Ser. A}\ ,\ \bibinfo
  {pages} {481--489}} (\bibinfo {year} {1961})}\BibitemShut {NoStop}%
\bibitem [{\citenamefont {de~Ruiter}\ \emph {et~al.}(2015)\citenamefont
  {de~Ruiter}, \citenamefont {Mugele},\ and\ \citenamefont {van~den
  Ende}}]{de2015air}%
  \BibitemOpen
  \bibfield  {author} {\bibinfo {author} {\bibfnamefont {J.}~\bibnamefont
  {de~Ruiter}}, \bibinfo {author} {\bibfnamefont {F.}~\bibnamefont {Mugele}}, \
  and\ \bibinfo {author} {\bibfnamefont {D.}~\bibnamefont {van~den Ende}},\
  }\bibfield  {title} {\enquote {\bibinfo {title} {Air cushioning in droplet
  impact. i. dynamics of thin films studied by dual wavelength reflection
  interference microscopy},}\ }\href {\doibase 10.1063/1.4906114} {\bibfield
  {journal} {\bibinfo  {journal} {Phys. Fluids}\ }\textbf {\bibinfo {volume}
  {27}},\ \bibinfo {pages} {012104} (\bibinfo {year} {2015})}\BibitemShut
  {NoStop}%
\bibitem [{\citenamefont {Derjaguin}(1943)}]{derjaguin1943}%
  \BibitemOpen
  \bibfield  {author} {\bibinfo {author} {\bibfnamefont {B.}~\bibnamefont
  {Derjaguin}},\ }\bibfield  {title} {\enquote {\bibinfo {title} {Thickness of
  liquid layer adhering to walls of vessels on their emptying and the theory of
  photo-and motion-picture film coating},}\ }\href@noop {} {\bibfield
  {journal} {\bibinfo  {journal} {Dokl. Acad. Sci. USSR}\ }\textbf {\bibinfo
  {volume} {39}},\ \bibinfo {pages} {13--16} (\bibinfo {year}
  {1943})}\BibitemShut {NoStop}%
\bibitem [{\citenamefont {Landau}\ and\ \citenamefont
  {Levich}(1942)}]{Landau1942}%
  \BibitemOpen
  \bibfield  {author} {\bibinfo {author} {\bibfnamefont {L.}~\bibnamefont
  {Landau}}\ and\ \bibinfo {author} {\bibfnamefont {V.}~\bibnamefont
  {Levich}},\ }\bibfield  {title} {\enquote {\bibinfo {title} {Dragging of a
  liquid by a moving plate},}\ }\href@noop {} {\bibfield  {journal} {\bibinfo
  {journal} {Acta Physicochim. USSR}\ }\textbf {\bibinfo {volume} {17}},\
  \bibinfo {pages} {42--54} (\bibinfo {year} {1942})}\BibitemShut {NoStop}%
\bibitem [{\citenamefont {Keiser}\ \emph {et~al.}(2017)\citenamefont {Keiser},
  \citenamefont {Keiser}, \citenamefont {Clanet},\ and\ \citenamefont
  {Qu{\'e}r{\'e}}}]{keiser2017drop}%
  \BibitemOpen
  \bibfield  {author} {\bibinfo {author} {\bibfnamefont {A.}~\bibnamefont
  {Keiser}}, \bibinfo {author} {\bibfnamefont {L.}~\bibnamefont {Keiser}},
  \bibinfo {author} {\bibfnamefont {C.}~\bibnamefont {Clanet}}, \ and\ \bibinfo
  {author} {\bibfnamefont {D.}~\bibnamefont {Qu{\'e}r{\'e}}},\ }\bibfield
  {title} {\enquote {\bibinfo {title} {Drop friction on liquid-infused
  materials},}\ }\href@noop {} {\bibfield  {journal} {\bibinfo  {journal} {Soft
  Matter}\ }\textbf {\bibinfo {volume} {13}},\ \bibinfo {pages} {6981--6987}
  (\bibinfo {year} {2017})}\BibitemShut {NoStop}%
\bibitem [{\citenamefont {Probstein}(2005)}]{probstein2005physicochemical}%
  \BibitemOpen
  \bibfield  {author} {\bibinfo {author} {\bibfnamefont {R.~F.}\ \bibnamefont
  {Probstein}},\ }\href@noop {} {\emph {\bibinfo {title} {Physicochemical
  hydrodynamics: an introduction}}}\ (\bibinfo  {publisher} {John Wiley \&
  Sons},\ \bibinfo {year} {2005})\BibitemShut {NoStop}%
\bibitem [{\citenamefont {Lhuissier}\ \emph {et~al.}(2013)\citenamefont
  {Lhuissier}, \citenamefont {Tagawa}, \citenamefont {Tran},\ and\
  \citenamefont {Sun}}]{lhuissier2013levitation}%
  \BibitemOpen
  \bibfield  {author} {\bibinfo {author} {\bibfnamefont {H.}~\bibnamefont
  {Lhuissier}}, \bibinfo {author} {\bibfnamefont {Y.}~\bibnamefont {Tagawa}},
  \bibinfo {author} {\bibfnamefont {T.}~\bibnamefont {Tran}}, \ and\ \bibinfo
  {author} {\bibfnamefont {C.}~\bibnamefont {Sun}},\ }\bibfield  {title}
  {\enquote {\bibinfo {title} {Levitation of a drop over a moving surface},}\
  }\href@noop {} {\bibfield  {journal} {\bibinfo  {journal} {J. Fluid Mech.}\
  }\textbf {\bibinfo {volume} {733}} (\bibinfo {year} {2013})}\BibitemShut
  {NoStop}%
\bibitem [{\citenamefont {van~der Veen}\ \emph {et~al.}(2012)\citenamefont
  {van~der Veen}, \citenamefont {Tran}, \citenamefont {Lohse},\ and\
  \citenamefont {Sun}}]{WLI}%
  \BibitemOpen
  \bibfield  {author} {\bibinfo {author} {\bibfnamefont {R.~C.~A.}\
  \bibnamefont {van~der Veen}}, \bibinfo {author} {\bibfnamefont
  {T.}~\bibnamefont {Tran}}, \bibinfo {author} {\bibfnamefont {D.}~\bibnamefont
  {Lohse}}, \ and\ \bibinfo {author} {\bibfnamefont {C.}~\bibnamefont {Sun}},\
  }\bibfield  {title} {\enquote {\bibinfo {title} {{Direct measurements of air
  layer profiles under impacting droplets using high-speed color
  interferometry}},}\ }\href {\doibase 10.1103/PhysRevE.85.026315} {\bibfield
  {journal} {\bibinfo  {journal} {Phys. Rev. E}\ }\textbf {\bibinfo {volume}
  {85}},\ \bibinfo {pages} {026315} (\bibinfo {year} {2012})},\ \Eprint
  {http://arxiv.org/abs/1111.3762} {1111.3762} \BibitemShut {NoStop}%
\bibitem [{\citenamefont {Park}\ \emph {et~al.}(2016)\citenamefont {Park},
  \citenamefont {Kim}, \citenamefont {Grinthal}, \citenamefont {He},
  \citenamefont {Fox}, \citenamefont {Weaver},\ and\ \citenamefont
  {Aizenberg}}]{park2016condensation}%
  \BibitemOpen
  \bibfield  {author} {\bibinfo {author} {\bibfnamefont {K.-C.}\ \bibnamefont
  {Park}}, \bibinfo {author} {\bibfnamefont {P.}~\bibnamefont {Kim}}, \bibinfo
  {author} {\bibfnamefont {A.}~\bibnamefont {Grinthal}}, \bibinfo {author}
  {\bibfnamefont {N.}~\bibnamefont {He}}, \bibinfo {author} {\bibfnamefont
  {D.}~\bibnamefont {Fox}}, \bibinfo {author} {\bibfnamefont {J.~C.}\
  \bibnamefont {Weaver}}, \ and\ \bibinfo {author} {\bibfnamefont
  {J.}~\bibnamefont {Aizenberg}},\ }\bibfield  {title} {\enquote {\bibinfo
  {title} {Condensation on slippery asymmetric bumps},}\ }\href@noop {}
  {\bibfield  {journal} {\bibinfo  {journal} {Nature (London)}\ }\textbf
  {\bibinfo {volume} {531}},\ \bibinfo {pages} {78--82} (\bibinfo {year}
  {2016})}\BibitemShut {NoStop}%
\bibitem [{\citenamefont {Spiers}\ \emph {et~al.}(1974)\citenamefont {Spiers},
  \citenamefont {Subbaraman},\ and\ \citenamefont
  {Wilkinson}}]{spiers_free_1974}%
  \BibitemOpen
  \bibfield  {author} {\bibinfo {author} {\bibfnamefont {R.~P.}\ \bibnamefont
  {Spiers}}, \bibinfo {author} {\bibfnamefont {C.~V.}\ \bibnamefont
  {Subbaraman}}, \ and\ \bibinfo {author} {\bibfnamefont {W.~L.}\ \bibnamefont
  {Wilkinson}},\ }\bibfield  {title} {\enquote {\bibinfo {title} {Free coating
  of a newtonian liquid onto a vertical surface},}\ }\href@noop {} {\bibfield
  {journal} {\bibinfo  {journal} {Chem. Eng. Sci.}\ }\textbf {\bibinfo {volume}
  {29}},\ \bibinfo {pages} {389--396} (\bibinfo {year} {1974})}\BibitemShut
  {NoStop}%
\bibitem [{\citenamefont {Bretherton}(1961)}]{bretherton_motion_1961}%
  \BibitemOpen
  \bibfield  {author} {\bibinfo {author} {\bibfnamefont {F.~P.}\ \bibnamefont
  {Bretherton}},\ }\bibfield  {title} {\enquote {\bibinfo {title} {The motion
  of long bubbles in tubes},}\ }\href@noop {} {\bibfield  {journal} {\bibinfo
  {journal} {J. Fluid Mech.}\ }\textbf {\bibinfo {volume} {10}},\ \bibinfo
  {pages} {166} (\bibinfo {year} {1961})}\BibitemShut {NoStop}%
\bibitem [{\citenamefont {White}\ and\ \citenamefont
  {Tallmadge}(1965)}]{white_theory_1965}%
  \BibitemOpen
  \bibfield  {author} {\bibinfo {author} {\bibfnamefont {D.~A.}\ \bibnamefont
  {White}}\ and\ \bibinfo {author} {\bibfnamefont {J.~A.}\ \bibnamefont
  {Tallmadge}},\ }\bibfield  {title} {\enquote {\bibinfo {title} {Theory of
  drag out of liquids on flat plates},}\ }\href@noop {} {\bibfield  {journal}
  {\bibinfo  {journal} {Chem. Eng. Sci.}\ }\textbf {\bibinfo {volume} {20}},\
  \bibinfo {pages} {33--37} (\bibinfo {year} {1965})}\BibitemShut {NoStop}%
\bibitem [{\citenamefont {Maleki}\ \emph {et~al.}(2011)\citenamefont {Maleki},
  \citenamefont {Reyssat}, \citenamefont {Restagno}, \citenamefont
  {Qu{\'e}r{\'e}},\ and\ \citenamefont {Clanet}}]{maleki_landaulevich_2011}%
  \BibitemOpen
  \bibfield  {author} {\bibinfo {author} {\bibfnamefont {M.}~\bibnamefont
  {Maleki}}, \bibinfo {author} {\bibfnamefont {M.}~\bibnamefont {Reyssat}},
  \bibinfo {author} {\bibfnamefont {F.}~\bibnamefont {Restagno}}, \bibinfo
  {author} {\bibfnamefont {D.}~\bibnamefont {Qu{\'e}r{\'e}}}, \ and\ \bibinfo
  {author} {\bibfnamefont {C.}~\bibnamefont {Clanet}},\ }\bibfield  {title}
  {\enquote {\bibinfo {title} {Landau-levich menisci},}\ }\href@noop {}
  {\bibfield  {journal} {\bibinfo  {journal} {J. Colloid Interface Sci.}\
  }\textbf {\bibinfo {volume} {354}},\ \bibinfo {pages} {359--363} (\bibinfo
  {year} {2011})}\BibitemShut {NoStop}%
\bibitem [{\citenamefont {Snoeijer}\ \emph {et~al.}(2006)\citenamefont
  {Snoeijer}, \citenamefont {Delon}, \citenamefont {Fermigier},\ and\
  \citenamefont {Andreotti}}]{snoeijer2006avoided}%
  \BibitemOpen
  \bibfield  {author} {\bibinfo {author} {\bibfnamefont {J.~H.}\ \bibnamefont
  {Snoeijer}}, \bibinfo {author} {\bibfnamefont {G.}~\bibnamefont {Delon}},
  \bibinfo {author} {\bibfnamefont {M.}~\bibnamefont {Fermigier}}, \ and\
  \bibinfo {author} {\bibfnamefont {B.}~\bibnamefont {Andreotti}},\ }\bibfield
  {title} {\enquote {\bibinfo {title} {Avoided critical behavior in dynamically
  forced wetting},}\ }\href@noop {} {\bibfield  {journal} {\bibinfo  {journal}
  {Phys. Rev. Lett.}\ }\textbf {\bibinfo {volume} {96}},\ \bibinfo {pages}
  {174504} (\bibinfo {year} {2006})}\BibitemShut {NoStop}%
\bibitem [{\citenamefont {Snoeijer}\ and\ \citenamefont
  {Andreotti}(2013)}]{snoeijer2013moving}%
  \BibitemOpen
  \bibfield  {author} {\bibinfo {author} {\bibfnamefont {J.~H.}\ \bibnamefont
  {Snoeijer}}\ and\ \bibinfo {author} {\bibfnamefont {B.}~\bibnamefont
  {Andreotti}},\ }\bibfield  {title} {\enquote {\bibinfo {title} {Moving
  contact lines: scales, regimes, and dynamical transitions},}\ }\href@noop {}
  {\bibfield  {journal} {\bibinfo  {journal} {Ann. Rev. Fluid Mech.}\ }\textbf
  {\bibinfo {volume} {45}} (\bibinfo {year} {2013})}\BibitemShut {NoStop}%
\bibitem [{\citenamefont {Urata}\ \emph {et~al.}(2015)\citenamefont {Urata},
  \citenamefont {Dunderdale}, \citenamefont {England},\ and\ \citenamefont
  {Hozumi}}]{urata2015self}%
  \BibitemOpen
  \bibfield  {author} {\bibinfo {author} {\bibfnamefont {C.}~\bibnamefont
  {Urata}}, \bibinfo {author} {\bibfnamefont {G.~J.}\ \bibnamefont
  {Dunderdale}}, \bibinfo {author} {\bibfnamefont {M.~W.}\ \bibnamefont
  {England}}, \ and\ \bibinfo {author} {\bibfnamefont {A.}~\bibnamefont
  {Hozumi}},\ }\bibfield  {title} {\enquote {\bibinfo {title} {Self-lubricating
  organogels (slugs) with exceptional syneresis-induced anti-sticking
  properties against viscous emulsions and ices},}\ }\href@noop {} {\bibfield
  {journal} {\bibinfo  {journal} {J. Mater. Chem. A}\ }\textbf {\bibinfo
  {volume} {3}},\ \bibinfo {pages} {12626--12630} (\bibinfo {year}
  {2015})}\BibitemShut {NoStop}%
\bibitem [{\citenamefont {Cui}\ \emph {et~al.}(2015)\citenamefont {Cui},
  \citenamefont {Daniel}, \citenamefont {Grinthal}, \citenamefont {Lin},\ and\
  \citenamefont {Aizenberg}}]{cui2015dynamic}%
  \BibitemOpen
  \bibfield  {author} {\bibinfo {author} {\bibfnamefont {J.}~\bibnamefont
  {Cui}}, \bibinfo {author} {\bibfnamefont {D.}~\bibnamefont {Daniel}},
  \bibinfo {author} {\bibfnamefont {A.}~\bibnamefont {Grinthal}}, \bibinfo
  {author} {\bibfnamefont {K.}~\bibnamefont {Lin}}, \ and\ \bibinfo {author}
  {\bibfnamefont {J.}~\bibnamefont {Aizenberg}},\ }\bibfield  {title} {\enquote
  {\bibinfo {title} {Dynamic polymer systems with self-regulated secretion for
  the control of surface properties and material healing},}\ }\href@noop {}
  {\bibfield  {journal} {\bibinfo  {journal} {Nat. Mat.}\ }\textbf {\bibinfo
  {volume} {14}},\ \bibinfo {pages} {790--795} (\bibinfo {year}
  {2015})}\BibitemShut {NoStop}%
\bibitem [{\citenamefont {Yuan}\ \emph {et~al.}(2008)\citenamefont {Yuan},
  \citenamefont {Liu}, \citenamefont {Akbulut}, \citenamefont {Hu},
  \citenamefont {Suib}, \citenamefont {Kong},\ and\ \citenamefont
  {Stellacci}}]{PDMSfunc}%
  \BibitemOpen
  \bibfield  {author} {\bibinfo {author} {\bibfnamefont {J.}~\bibnamefont
  {Yuan}}, \bibinfo {author} {\bibfnamefont {X.}~\bibnamefont {Liu}}, \bibinfo
  {author} {\bibfnamefont {O.}~\bibnamefont {Akbulut}}, \bibinfo {author}
  {\bibfnamefont {J.}~\bibnamefont {Hu}}, \bibinfo {author} {\bibfnamefont
  {S.~L.}\ \bibnamefont {Suib}}, \bibinfo {author} {\bibfnamefont
  {J.}~\bibnamefont {Kong}}, \ and\ \bibinfo {author} {\bibfnamefont
  {F.}~\bibnamefont {Stellacci}},\ }\bibfield  {title} {\enquote {\bibinfo
  {title} {Superwetting nanowire membranes for selective absorption},}\
  }\href@noop {} {\bibfield  {journal} {\bibinfo  {journal} {Nat.
  Nanotechnol.}\ }\textbf {\bibinfo {volume} {3}},\ \bibinfo {pages} {332--336}
  (\bibinfo {year} {2008})}\BibitemShut {NoStop}%
\bibitem [{\citenamefont {B.}(2017)}]{tracker}%
  \BibitemOpen
  \bibfield  {author} {\bibinfo {author} {\bibfnamefont {Douglas}\ \bibnamefont
  {B.}},\ }\href@noop {} {\enquote {\bibinfo {title} {Tracker video analysis
  and modeling tool},}\ }\bibinfo {howpublished}
  {\url{http://physlets.org/tracker/}} (\bibinfo {year} {2017}),\ \bibinfo
  {note} {accessed: 2017-07-30}\BibitemShut {NoStop}%
\end{thebibliography}

\providecommand{\noopsort}[1]{}\providecommand{\singleletter}[1]{#1}%

\end{document}